\documentclass[a4paper,11pt]{article}
\usepackage{jcappub}
\usepackage{makecell}
\usepackage{graphicx}
\usepackage{hyperref}
\newcommand{\angstrom}{\textup{\AA}}
\newcommand{\lya}{Ly$\alpha$~}
\newcommand{\al}{$\alpha$}

\usepackage{graphicx}
\usepackage{amsmath, makecell, booktabs, enumitem}
\usepackage{orcidlink}

\title{\boldmath Probing the limits of cosmological information from the Lyman-\al~forest 2-point correlation functions}

\affiliation{Affiliations are in Appendix \ref{sec:affiliations}}
\emailAdd{turner.1839@osu.edu}

\author[a,b]{{Wynne~Turner}\orcidlink{0009-0008-3418-5599},}
\author[c]{{Andrei~Cuceu}\orcidlink{0000-0002-2169-0595},\footnote{NASA Einstein Fellow}}
\author[a,b,d]{{Paul~Martini}\orcidlink{0000-0002-4279-4182},}

\author[c]{{J.~Aguilar},}
\author[e]{{S.~Ahlen}\orcidlink{0000-0001-6098-7247},}
\author[c]{{A.~Anand}\orcidlink{0000-0003-2923-1585},}
\author[f,g]{{D.~Bianchi}\orcidlink{0000-0001-9712-0006},}
\author[h]{{D.~Brooks},}
\author[i]{{L.~Casas},}
\author[c]{{T.~Claybaugh},}
\author[j]{{A.~de la Macorra}\orcidlink{0000-0002-1769-1640},}
\author[k,l]{{B.~Dey}\orcidlink{0000-0002-5665-7912},}
\author[h]{{P.~Doel},}
\author[c,m]{{S.~Ferraro}\orcidlink{0000-0003-4992-7854},}
\author[i]{{A.~Font-Ribera}\orcidlink{0000-0002-3033-7312},}
\author[n,o]{{J.~E.~Forero-Romero}\orcidlink{0000-0002-2890-3725},}
\author[p,q,r]{{E.~Gazta\~naga}\orcidlink{0000-0001-9632-0815},}
\author[c,s]{{S.~Gontcho A Gontcho}\orcidlink{0000-0003-3142-233X},}
\author[t]{{G.~Gutierrez},}
\author[u,v]{{H.~K.~Herrera-Alcantar}\orcidlink{0000-0002-9136-9609},}
\author[a,d,w]{{K.~Honscheid}\orcidlink{0000-0002-6550-2023},}
\author[x]{{M.~Ishak}\orcidlink{0000-0002-6024-466X},}
\author[y]{{R.~Joyce}\orcidlink{0000-0003-0201-5241},}
\author[z]{{R.~Kehoe},}
\author[aa]{{D.~Kirkby}\orcidlink{0000-0002-8828-5463},}
\author[c]{{A.~Kremin}\orcidlink{0000-0001-6356-7424},}
\author[h]{{O.~Lahav},}
\author[c]{{M.~Landriau}\orcidlink{0000-0003-1838-8528},}
\author[ab]{{L.~Le~Guillou}\orcidlink{0000-0001-7178-8868},}
\author[ac,i]{{M.~Manera}\orcidlink{0000-0003-4962-8934},}
\author[ad,i]{{R.~Miquel},}
\author[j]{{A.~Mu\~noz-Guti\'errez},}
\author[q]{{S.~Nadathur}\orcidlink{0000-0001-9070-3102},}
\author[ae,af]{{G.~Niz}\orcidlink{0000-0002-1544-8946},}
\author[v,c]{{N.~Palanque-Delabrouille}\orcidlink{0000-0003-3188-784X},}
\author[ag,ah,ai]{{W.~J.~Percival}\orcidlink{0000-0002-0644-5727},}
\author[c,aj,m]{{C.~Poppett},}
\author[ak]{{F.~Prada}\orcidlink{0000-0001-7145-8674},}
\author[a,b]{{A.~J.~Ross}\orcidlink{0000-0002-7522-9083},}
\author[al]{{G.~Rossi},}
\author[am]{{E.~Sanchez}\orcidlink{0000-0002-9646-8198},}
\author[c]{{D.~Schlegel},}
\author[an,ao]{{M.~Schubnell},}
\author[ap]{{H.~Seo}\orcidlink{0000-0002-6588-3508},}
\author[c]{{J.~Silber}\orcidlink{0000-0002-3461-0320},}
\author[y]{{D.~Sprayberry},}
\author[ao]{{G.~Tarl\'{e}}\orcidlink{0000-0003-1704-0781},}
\author[aq,ar]{{M.~Walther}\orcidlink{0000-0002-1748-3745},}
\author[y]{{B.~A.~Weaver},}
\author[c]{{R.~Zhou}\orcidlink{0000-0001-5381-4372},}
\author[as]{{H.~Zou}\orcidlink{0000-0002-6684-3997}}

\abstract{
The standard cosmological analysis with the \lya forest relies on a continuum fitting procedure that suppresses information on large scales and distorts the three-dimensional correlation function on all scales. In this work, we present the first cosmological forecasts without continuum fitting distortion in the \lya forest, focusing on the recovery of large-scale information. Using idealized synthetic data, we compare the constraining power of the full shape of the \lya forest auto-correlation and its cross-correlation with quasars using the baseline continuum fitting analysis versus the true continuum. We find that knowledge of the true continuum enables a $\sim10\%$ reduction in uncertainties on the Alcock-Paczyński (AP) parameter and the matter density, $\Omega_\mathrm{m}$. We also explore the impact of large-scale information by extending the analysis up to separations of $240\,h^{-1}\mathrm{Mpc}$ along and across the line of sight. The combination of these analysis choices can recover significant large-scale information, yielding up to a $\sim15\%$ improvement in AP constraints. This improvement is analogous to extending the \lya forest survey area by $\sim40\%$.}

\begin{document}
\maketitle
\flushbottom

\section{Introduction} \label{sec:intro}

Ever since the discovery of the universe's accelerating expansion \citep{riess_1998_supernovae,perlmutter_1999_supernovae}, determining the nature of dark energy has become a central goal in modern cosmology. Multiple independent probes -- Type 1a supernovae, the cosmic microwave background, and the clustering of tracers of the matter-density field -- consistently confirm this late-time accelerated expansion. Among the most powerful tools for constraining the expansion history are baryon acoustic oscillations (BAO), which provide a standard ruler originating from oscillations in the photon-baryon plasma prior to recombination. While galaxies and quasars are often used as tracers of the matter-density field \citep{2df,sdss_bao_2005} in which we can measure the BAO scale, their number densities decline with increasing redshift, necessitating alternative high-redshift probes.

The Lyman-alpha (Ly\al) forest is the collection of absorption lines seen in the spectra of distant quasars resulting from intervening neutral hydrogen along the line of sight. At redshifts $z>2$, the \lya forest is observable from the ground and becomes a valuable probe for cosmological measurements at high redshifts, as the BAO scale can be measured from the distribution of matter in the intergalactic medium. The Baryon Oscillation Spectroscopic Survey \citep[BOSS;][]{dawson_boss_2013}, as part of the third stage of the Sloan Digital Sky Survey \citep[SDSS-III;][]{eisenstein_sdss3_2011} was the first to obtain cosmological measurements from the BAO scale in the \lya forest \citep[e.g.][]{busca_2013_lya_boss,slosar_2013_lya_boss,fontribera_2014_lyaqso_boss}. This was repeated by its successor, extended-BOSS \citep[eBOSS;][]{dawson_eboss_2016} within SDSS-IV \citep[][]{blanton_sdss4_2017} with a sample of $\sim$ 250,000 \lya forest quasars \citep[e.g.][]{ahumada_eboss_2020,bourboux_completed_2020}.

The Dark Energy Spectroscopic Instrument \citep[DESI;][]{desi_collaboration_desi_2016,desi_instrument_2016,miller_optical_2024,poppett_fiber_2024} is a Stage-IV spectroscopic survey operating on the Mayall 4-meter telescope at Kitt Peak National Observatory \citep{desi_instrument_overview_2022}. DESI is on schedule to measure the spectra of approximately 50 million galaxies and quasars within the first five years of operation \cite{desi_y1_dr1}. Approximately one million of these targets are high-redshift ($z\geq2.1$) quasars whose spectra include the \lya forest. DESI selects targets with a sophisticated pipeline for survey tiling and optimization \citep{schlafly_surveyops_2023}. The observations are then processed by the spectroscopic pipeline \citep{guy_data_pipeline_2023} for further analysis. The main DESI survey began in May 2021 following successful commissioning and survey validation periods \citep{desi_collaboration_validation_2023}. Data Release 1 \citep[DR1;][]{desi_y1_dr1} is publicly available, containing data from the first year of the survey, alongside their cosmological interpretation \citep[e.g.][]{desi_y1_7}. Recently, cosmological results from the first three years of data (DR2) were also released, based on nearly one million \lya forest quasar spectra \citep{desi_dr2_lya} and over 14 million lower redshift galaxies and quasars \citep{desi_dr2_discretetracers}. The results have revealed tension between the best-fit model describing the universe's expansion history and the $\Lambda$CDM concordance model, instead favoring an evolving $w_0w_a$CDM model. In addition to the collection of more data, new analysis methods will be crucial in elucidating this tension. In particular, \cite{cuceu_2025} highlights the potential of the \lya forest to constrain dynamical dark energy at high redshift.

Most cosmological studies with the \lya forest have only used measurements of the BAO peak position. While robust, this method discards additional information contained in the full shape of the correlation functions. When the fiducial cosmology used to convert observed angular and redshift separations into comoving distances differs from the true cosmology, this introduces an anisotropy in the measured correlation functions. This anisotropy, known as the Alcock-Paczyński \citep[AP;][]{AP_1979} effect, can be extracted to provide even more precise cosmological constraints. It impacts both the BAO scale and the broadband of the correlation functions. A full-shape analysis with the \lya forest which extracts information from the broadband in addition to the BAO scale can provide up to twice the constraining power as a BAO-only analysis \citep{cuceu_cosmology_2021,cuceu23a,cuceu23b}. In a full-shape analysis, \cite{gerardi_2023} showed that most of the large-scale cosmological information is captured in compressed parameters that describe the BAO scale alongside anisotropies introduced by the AP effect and redshift space distortions (RSD). The \lya RSD signal is sensitive to the product of the linear growth rate and the amplitude of matter fluctuations on $8\,h^{-1}\,\mathrm{Mpc}$ scales, $f\sigma_8$, but is degenerate with an unknown velocity divergence bias \citep[e.g.][]{mcdonald_2003,slosar_2011,seljak_2012,givans_hirata_2020,chen_vlah_white_2021}. However, $f\sigma_8$ can be measured from the quasar RSD signal with the Ly\al-QSO cross-correlation. While the joint full-shape \lya forest analysis allows for a measurement of $f\sigma_8$, its most powerful cosmological constraint comes from the AP signal \citep[e.g.][]{cuceu_cosmology_2021}. A full-shape analysis with the \lya forest has recently been performed for the first time with DESI DR1 data, yielding a 1.6\% constraint on the AP effect \cite{cuceu_2025}.

Multiple systematics complicate the \lya forest full-shape analysis. One such systematic is the quasar continuum fitting method. The standard approach estimates a mean continuum from the dataset and performs a power-law fit to each forest. This process biases the mean and slope of each forest to zero, projecting out large-scale modes and correlating all pixels in the forest. Variations in quasar diversity become coupled with large-scale over- or under-densities during continuum fitting, leading to a distortion of the correlation function on all scales. The current solution is to model this effect with a distortion matrix prior to fitting the model to the measured correlation functions \citep{bautista_2017,busca_distortion_matrix}. The distortion matrix relies on several assumptions that must be validated using mock datasets \citep{busca_distortion_matrix}. So far, validation has focused on the BAO and AP parameters in the regimes typically studied \citep[e.g.][]{cuceu23b,casas_2025_validation,cuceu_2025}, but it has not yet been extended to other parameters. We do not study potential biases induced by the distortion matrix and instead focus on the idealized case assuming perfect knowledge of this distortion. Of particular importance to this work is that the continuum fitting process removes information about large-scale over- or under-densities that could in principle be recovered if the true unabsorbed continuum were known. Extending the analysis to larger scales may further enhance the recovery of this information. 

Recently, alternative continuum fitting approaches have shown promise in accessing this previously lost information. These include principal component analysis \citep[PCA; e.g.][]{suzuki_predicting_2005, paris_principal_2011, lee_mean-flux_2012}, nonnegative matrix factorization \citep[NMF; e.g.][]{zhu_nonnegative_2016}, and more sophisticated machine learning methods \citep{liu_deep_2021,sun_quasar_2023,turner_2024,hahn_spenderq_2025,pistis_continuum_DL_2025}. Approaches such as Quasar Factor Analysis \citep{sun_quasar_2023} and \textsc{SpenderQ} \citep{hahn_spenderq_2025} use both the red side and the blue side of the \lya emission line to estimate the unabsorbed forest continuum. More recently, neural network approaches \citep[e.g.][]{liu_deep_2021,turner_2024,pistis_continuum_DL_2025} predict the unabsorbed continuum independently of the forest region, thereby avoiding the coupling between large-scale structure and quasar diversity that drives distortions in methods that use data in the forest region. However, such predictions are not necessarily free of spurious correlations, for example from correlated continuum errors, and it remains important to test for and characterize these effects. Predicting the continuum independently of the forest also offers other advantages; for instance, LyCAN has been applied to DESI DR1 data to measure the evolution of the effective optical depth \citep{turner_2024}. PCA-based approaches, along with the neural networks iQNet \citep{liu_deep_2021} and LyCAN \citep{turner_2024}, have published predictions of the forest continuum based solely on the red side. Of these, the latter was developed to achieve accurate performance on DESI-like spectra. Tools such as LyCAN offer a path forward toward unlocking the full constraining power of the \lya forest.

The current state-of-the-art analysis \citep[e.g.][]{desi_y1_4, desi_dr2_lya} fits a model to the two-point correlation functions over separations from $r=30-180\,h^{-1}\,\mathrm{Mpc}$, and estimates the covariance matrix from the data by subsampling correlation function measurements from $\sim1000$ individual sky regions \cite{desi_dr2_lya}. While modeling small-scale correlations ($r_\mathrm{min}<30\,h^{-1}\,\mathrm{Mpc}$) is challenging due to nonlinearities, extending the analysis to larger scales may be more feasible. Full-shape analyses can access additional cosmological information at these larger scales, particularly from the AP effect. However, a significant challenge arises from the fact that the correlation functions are measured in two dimensions, resulting in a large $p\times p$ covariance matrix, where $p$ is the number of correlation function bins. As $p$ increases, it becomes increasingly difficult to accurately estimate the covariance matrix from the data alone.

The goal of this work is to investigate potential gains in cosmological constraining power through two key avenues: (1) assuming knowledge of the true continuum, which eliminates the distortion introduced by continuum fitting and is expected to better preserve large-scale information, and (2) extracting additional information by extending the analysis to larger scales. Our forecasts focus on the idealized case of perfect continuum knowledge to probe the limit of how much information could be gained with access to the unabsorbed continuum. Throughout this work, we refer to analyses that use the true continuum as \textit{undistorted} methods, as they do not require the distortion matrix. We assess improvements in both compressed and uncompressed (direct) cosmological parameters. Our aim is to quantify the additional constraining power achievable with the idealized undistorted analysis relative to the continuum fitting (\textit{distorted}) method, as well as the benefit of recovering large-scale information in both analyses.

This paper is organized as follows: we describe the synthetic data we use and different ways of estimating the quasar continuum in Section~\ref{sec:continuum}. In Section~\ref{sec:method} we describe how we estimate covariance matrices and our forecasting methodology. We present our cosmological forecasts on compressed parameters in Section~\ref{sec:results_compressed}, and on direct cosmological parameters in Section~\ref{sec:results_direct}. Throughout this paper we assume the flat $\Lambda$CDM cosmology from Planck 2018 \cite{planck_2018_cosmology}. In Section~\ref{sec:discussion} we provide a discussion including future work, and we summarize our results in Section~\ref{sec:summary}.

\section{Quasar Spectra}\label{sec:continuum}

In this section we describe the synthetic quasar spectra that are the foundation of this analysis, as well as how the quasar continuum is modeled. We first describe the synthetic data used for our forecasts in Section~\ref{sec:synthdata}. In Section~\ref{sec:forestcontinuum} we summarize the continuum fitting procedure employed in the DESI \lya forest analysis and compare it to the true unabsorbed continuum. 

\subsection{Synthetic Data}\label{sec:synthdata}

We use 50 independent synthetic realizations (mocks) of the first three years of DESI to compute covariance matrices. The mocks were generated by \texttt{quickquasars}\footnote{\url{https://github.com/desihub/desisim/blob/main/py/desisim/scripts/quickquasars.py}}, a DESI software suite that simulates realistic quasar spectra and is described in \cite{herrera-alcantar_synthetic_2023}. These mocks include input quasar continua produced by \texttt{simqso}\footnote{\url{https://github.com/imcgreer/simqso}}, a package that generates mock quasar spectra and photometry \cite{simqso}. This package models the continuum as a broken power-law and is optimized to better match the mean continuum observed in eBOSS DR16 \citep{bourboux_completed_2020} using principal component analysis (PCA). These unabsorbed continua are then multiplied by transmission files generated with \texttt{LyaCoLoRe}, a package that produces synthetic \lya forest datasets \citep{farr_lyacolore_2020}.

The \texttt{LyaCoLoRe} mocks were recently updated to include more realistic small-scale clustering and broadening of the BAO peak due to the nonlinear growth of structure \cite{casas_2025_validation}. We use these improved mocks for this work. Each mock is representative of the DESI DR2 \lya forest dataset with realistic noise and a comparable magnitude and redshift distribution. For more details on the DESI mock creation process, see \cite{herrera-alcantar_synthetic_2023} and \cite{casas_2025_validation}. For this work, we use mocks without contamination from metals, broad absorption lines (BALs), or optically thick absorbers to isolate the effects of the distortion introduced by continuum fitting. We use the true continua of the mocks for the undistorted analysis, and compare this to the continuum fitting (distorted) analysis to quantify the improvement in constraining power enabled by perfect knowledge of the true continuum. The computation of the covariance matrix is described in Section~\ref{sec:covariance}.

\subsection{Forest Continuum}\label{sec:forestcontinuum}

The current state-of-the-art in \lya forest cosmology involves fitting the expected flux in each forest in order to measure the amount of absorption present in each pixel. While this approach effectively captures the BAO signal, it removes information on scales comparable to the forest length, limiting the ability to fully exploit the full shape of the 3D correlation functions.

\begin{figure}
    \centering
    \includegraphics[width=\linewidth]{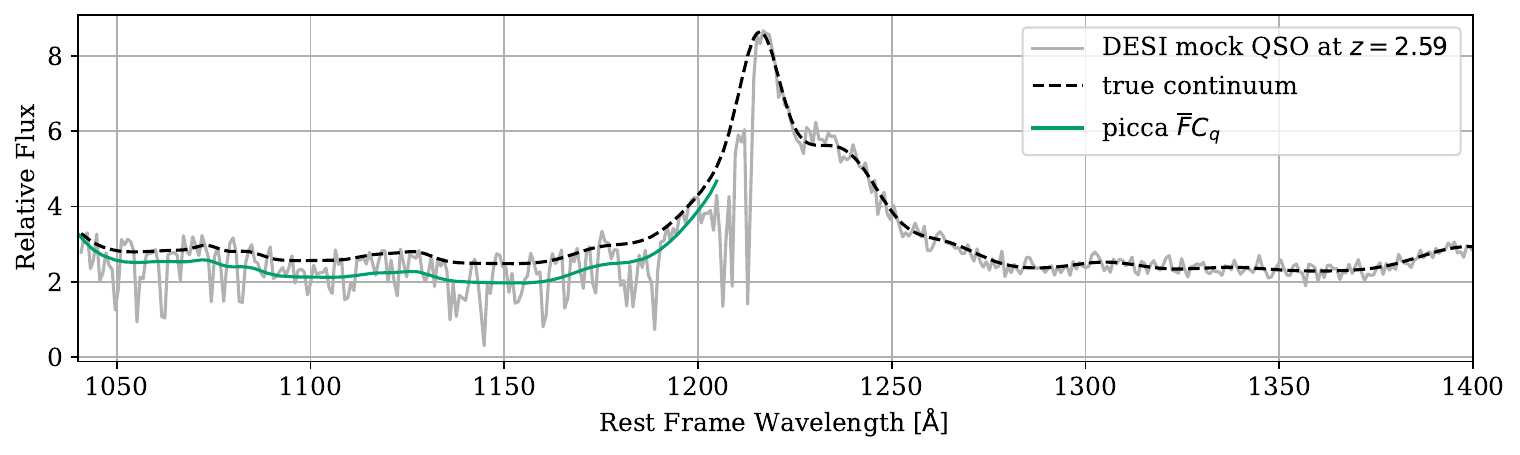}
    \caption{An example DESI DR2 mock quasar spectrum with its true continuum (\textit{black dashed line}) and the product $\overline{F}C_q$ fit by the \texttt{picca} pipeline \textit{(teal line)}.}
    \label{fig:lycan_spec}
\end{figure}

The DESI \lya forest analysis pipeline, \texttt{picca}, is a package that performs continuum fitting, measures correlation functions and power spectra, and computes the covariance matrices \cite{bourboux_completed_2020}. During continuum fitting, the expected flux $\overline{F}(\lambda)C_q(\lambda)$ is estimated by fitting a two-parameter model to the forest region of each quasar \citep{ramirez_lya_edr}:
\begin{equation}
\overline{F}(\lambda)C_q(\lambda) = \overline{C}(\lambda_{\rm RF})\left(a_q + b_q\frac{\Lambda-\Lambda_{\rm min}}{\Lambda_{\rm max}-\Lambda_{\rm min}}\right),
\end{equation}
where $\Lambda\equiv\log\lambda$, $\overline{C}(\lambda_{\rm RF})$ is the mean continuum measured from the data, $C_q(\lambda)$ is the unknown unabsorbed continuum, and $\overline{F}(\lambda)$ is the mean flux. The parameters $a_q$ and $b_q$ account for variations in quasar diversity and the redshift evolution of the mean flux. This procedure removes information on large scales, since fluctuations in the density field on scales comparable to the forest are absorbed into the fit parameters. It also introduces spurious correlations that affect all scales, which must be taken into account using a distortion matrix. The distortion matrices for the auto- and cross-correlation are defined as \cite{bautista_2017, dmdb_2017, busca_distortion_matrix}:
\begin{equation}
    D_{AB}^\mathrm{auto}=\frac{1}{W_A}\sum_{i,j\in A}w_iw_j\sum_{i',j'\in B}\eta_{ii'}\eta_{jj'},
\end{equation}
\begin{equation}
    D_{AB}^\mathrm{cross}=\frac{1}{W_A}\sum_{i,j\in A}w_iw_j\sum_{i',j\in B}\eta_{ii'},
\end{equation}
where $w_{i,j}$ are the weights, $A$ and $B$ refer to the data and model bins, respectively, and the sums are over pixel-pixel pairs for the auto-correlation and pixel-quasar pairs for the cross-correlation. The projection matrices, $\eta_{ij}$, are given by:
\begin{equation}
    \eta_{ij}=\delta_{ij}^K-\frac{w_j}{\sum_kw_k}-\frac{w_j\kappa_i\kappa_j}{\sum_kw_k\kappa_k^2},
\end{equation}
where $\delta_{ij}^K$ is the Kronecker delta. The quantity $\kappa_k=\log\lambda_k-\overline{\log\lambda_q}$, where $\lambda_i$ is the observed wavelength in pixel $i$ and $\overline{\log\lambda_q}$ is the mean of $\log\lambda$ in the forest of quasar $q$. The measured distortion matrix is multiplied by the undistorted correlation function model to achieve a distorted correlation function model, which is subsequently fit to the data. This method has been shown to yield robust, unbiased constraints for BAO and AP parameters \citep[see e.g.][]{cuceu23b,casas_2025_validation,cuceu_2025}, but its performance for other parameters has not yet been established \citep{busca_distortion_matrix}. Figure~\ref{fig:lycan_spec} shows an example DESI mock spectrum with the fitted continuum (expected flux) and its true continuum. Continuum prediction models such as LyCAN \cite{turner_2024} aim to reconstruct this true continuum independently of the forest region to achieve undistorted correlation function measurements. Recent results from LyCAN, presented in Appendix~\ref{sec:appendix_lycan}, indicate this may be feasible and therefore motivate this study.

In this work, we present the first analysis of the cosmological impact of bypassing the distortion matrix and preserving all of the large-scale information. To forecast the limiting case of perfect knowledge of the true continuum, we use the true continua from our mock dataset. 

\section{Method}\label{sec:method}

We perform cosmological forecasts using a noiseless synthetic data vector built from the fiducial cosmology \citep{planck_2018_cosmology} and DESI DR2 mock covariance matrices for the true continuum and continuum fitting analyses. Section~\ref{sec:covariance} describes how we compute mock covariances for the undistorted and distorted analyses. We provide an overview of the analysis framework in Section~\ref{sec:framework}, and in Section~\ref{sec:implementation} we describe how we perform our forecasts in two different parameter spaces to constrain compressed (\S\ref{sec:compressed}) or direct (\S\ref{sec:directfit}) cosmological parameters.

\subsection{Correlations and Covariance}\label{sec:covariance}

We use two versions of synthetic 3D \lya correlation functions without contaminants. The distorted version includes a model for the distortion effect due to continuum fitting and a distortion matrix computed from one mock, while the undistorted version is based on the true continuum and does not require the distortion matrix. In the distorted analysis, the model for the correlation function is multiplied by the distortion matrix. We generate our synthetic data vector following the framework of \cite{cuceu_cosmology_2021}, built from a model based on cosmological parameters from the Planck 2018 results \citep{planck_2018_cosmology} and nuisance parameters from DESI DR2 \citep{desi_dr2_lya}. The data vector is noiseless and we calculate covariance matrices from uncontaminated synthetic data based on the first three years of DESI.

\begin{figure}
    \centering
    \includegraphics[width=\linewidth]{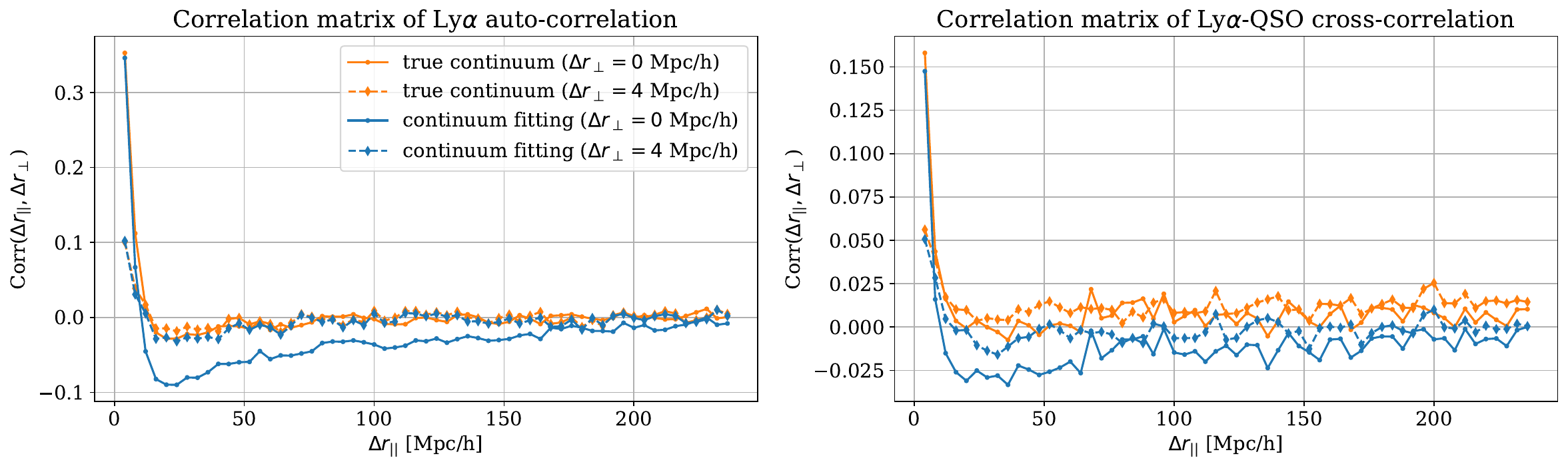}
    \caption{The unsmoothed correlation matrix as a function of line-of-sight separation for the true continuum (undistorted) and continuum fitting (distorted) analyses. We show correlation matrix elements for the \lya auto-correlation (\textit{left}) and the Ly$\alpha$-QSO cross-correlation (\textit{right}) in two bins of transverse separation. The correlation quickly approaches zero with increasing separation, especially for the true continuum case which exhibits smaller correlations.}
    \label{fig:corrmats}
\end{figure}

We use \texttt{picca} to compute correlation functions and synthetic covariance matrices from the mock data described in Section~\ref{sec:synthdata}. We only use the covariance matrices since we are only concerned with forecasting, while the noiseless synthetic data used in our forecasts is generated from the fiducial model described in Section~\ref{sec:framework}. In the baseline continuum fitting analysis, the covariance matrix is normally computed through subsampling of HEALPix pixels \citep{healpix_2005} with \texttt{NSIDE = 16}. This assumes that the correlation function in each $\sim250\times250\,(  h^{-1} \rm Mpc)^2$ patch covered by one HEALPix pixel is independent. While this is a good assumption for the continuum fitting analysis where large-scale modes are projected out and the fits only extend to 180 Mpc/$h$, it may no longer be accurate for an undistorted continuum analysis where we expect large-scale correlations. There are therefore two main considerations when deciding how to compute the correlations and covariance: (1) the size of the HEALPix pixels relative the largest transverse separations, and (2) the number of independent correlation function measurements that we use to derive the covariance. To ensure independence at larger scales, we use larger HEALPix pixels (\texttt{NSIDE = 14}, corresponding to patches of linear size $\sim280\, h^{-1} \rm Mpc$) to compute the subsamples for the true continuum analysis. Since the larger HEALPix pixels reduce the number of available subsamples for estimating the covariance matrix, we mitigate this by using 50 independent mock realizations, providing a sufficient number of independent correlation function measurements ($N=N_\mathrm{HEALPix}\times N_\mathrm{mocks}\approx43000$) relative to the size of the data vector. We compute the covariance up to separations of $240\, h^{-1} \rm Mpc$ in both the transverse and line-of-sight directions to investigate the additional constraining power that may come from larger scales, particularly for the undistorted analysis. For consistency, we compute the covariance matrix for the distorted analysis in the same way. We rescale each covariance matrix to one mock by multiplying them by a factor of 50. We also calculate the cross-covariance between the \lya auto- and cross-correlation functions as in \cite{desi_y1_4} and \cite{desi_dr2_lya}. Figure~\ref{fig:corrmats} shows representations of the correlation matrix for each analysis for the \lya auto- and cross-correlation functions. The variance of the undistorted correlation function is about $5-10\%$ larger than that of the distorted correlation function in the line-of-sight bin ($\Delta r_\perp=0~\mathrm{Mpc}/h$).

In the baseline continuum fitting analysis, the noisy measurement of the covariance matrix from the data is smoothed. This process replaces all off-diagonal elements of the correlation matrix with equal line-of-sight and transverse separations with their average \cite[e.g.][]{delubac_boss_dr11, bautista_2017, bourboux_completed_2020, cuceu_desi_lya_validation, desi_y1_4, desi_dr2_lya}. While the smoothing procedure has been validated in the context of continuum fitting for BAO and AP parameters \cite{cuceu23b,cuceu_desi_lya_validation,cuceu_2025}, it has not been validated in the true continuum analysis or for other parameters of interest. We choose not to adopt the assumptions implicit in the smoothing procedure, as they may not hold in the undistorted analysis. They also may not hold at larger separations as the smoothing erases some large-scale structure in the correlation matrix (e.g. due to cosmic variance). For these reasons, we do not smooth our covariance matrices in either analysis, except to compare our results to the current baseline.

Because we use a limited number of independent samples to measure covariance matrices, these estimates will be noisy, particularly at larger separations. Noise in the covariance matrix tends to result in underestimated uncertainties in inferred parameters due to the bias in the inverse covariance \citep{hartlap_2007}. Since we do not smooth our covariance matrices, we apply the Hartlap correction \citep{hartlap_2007, percival_covariance_2014} to the inverse covariance during the likelihood analysis. We validate that the Hartlap approximation is accurate and sufficient in the regimes we explore in Appendix~\ref{sec:appendix_hartlap}.

This work uses the standard estimators of the correlation function and its associated covariance matrix currently employed in the DESI \lya forest analysis \citep{desi_y1_4,desi_dr2_lya}. However, these current estimators are sub-optimal. During correlation function estimation, pixel weights are assigned independently without accounting for covariance between pixels within the same forest (cross-pixel) or between different forests (cross-forest) \citep[e.g.][]{slosar_2013_lya_boss,ramirez_lya_edr}, whereas an optimal estimator would take into account the full covariance matrix of all pixels in the survey. The size of the \lya forest dataset precludes the use of an optimal estimator at present. Instead, \cite{ramirez_lya_edr} introduced an empirical correction term tuned on mock datasets that improves the precision of the correlation function measurements. In the context of the covariance matrix, the subsampling estimator is also sub-optimal. While these estimators have been extensively validated for the traditional continuum fitting analysis, their performance in true continuum analyses remains relatively untested. Developing a more optimal covariance estimator tailored for undistorted continuum analyses will be an important goal for future work. 

\subsection{Theoretical Framework}\label{sec:framework}

The correlation functions are computed on grids of comoving coordinates along and across the line of sight. For two tracers at redshifts $z_i$ and $z_j$ separated by an angle $\Delta\theta$, their separations in comoving coordinates are \citep{desainte_eboss_2019,bourboux_completed_2020,cuceu_cosmology_2021}:
\begin{equation}\label{eqn:rpar}
    r_{\parallel} = [D_{\rm C,fid}(z_i)-D_{\rm C,fid}(z_j)]\cos(\Delta\theta/2),
\end{equation}
\begin{equation}\label{eqn:rperp}
    r_{\perp}=[D_{\rm M,fid}(z_i)+D_{\rm M,fid}(z_j)]\sin(\Delta\theta/2),
\end{equation}
where $D_C=c\int_0^z dz'/H(z')$ is the radial comoving distance, $D_M$ is the transverse comoving distance, and the $\textrm{fid}$ subscript denotes the fiducial cosmology. Throughout this work, we adopt the Planck 2018 $\Lambda$CDM model as our fiducial cosmology \citep{planck_2018_cosmology}. The mismatch between the fiducial and true cosmology induces an extra anisotropy in the correlation functions due to the AP effect, which we extract for additional constraining power.

The tracer separations $r_\parallel$ and $r_\perp$ correspond to the separations computed in the fiducial cosmology. If this differs from the true cosmology, the true separations are instead described by transformations of the fiducial separations, $r_\parallel'(r_\parallel,r_\perp,z)$ and $r_\perp'(r_\parallel,r_\perp,z)$ \citep{kirkby_2013, cuceu_cosmology_2021}. Since the AP effect changes the ratio $r_\perp/r_\parallel$, these true separations can be related to the fiducial ones through anisotropic rescalings,
\begin{equation}
    r_\parallel' = q_\parallel r_\parallel,
\end{equation}
\begin{equation}
    r_\perp' = q_\perp r_\perp.
\end{equation}
The AP parameter at effective redshift $z$ is defined as the ratio of the transverse comoving distance $D_M(z)$ to the Hubble distance $D_H(z)$:
\begin{equation}
    F_{\rm AP} = \frac{D_M(z)}{D_H(z)}.
\end{equation}
To isolate the anisotropy introduced by the AP effect, we measure the parameter
\begin{equation}\label{eqn:phi}
    \phi(z) = \frac{F_{\rm AP}(z)}{F_{\rm AP,fid}(z)} = \frac{q_\perp(z)}{q_\parallel(z)} = \frac{D_M(z)H(z)}{[D_M(z)H(z)]_{\rm fid}}.
\end{equation}
A measurement of $\phi(z)$ therefore corresponds to a measurement of $\Omega_\mathrm{m}$ in flat $\Lambda$CDM \citep{cuceu_cosmology_2021}. This information is commonly summarized as $\phi_f$, which captures the full-shape AP signal. Alternatively, the AP information can be separated into two components: $\phi_p$, which isolates the contribution from the BAO peak, and $\phi_s$, which captures the broadband (smooth) contribution. Similarly, we can define an isotropic rescaling of coordinates as
\begin{equation}\label{eqn:alpha}
    \alpha(z) = \sqrt{q_\perp(z)q_\parallel(z)}.
\end{equation}
The compressed parameters $\alpha(z)$ and $\phi(z)$ will both be unity if the true underlying cosmology is equal to the fiducial cosmology. The parameter $\alpha(z)$ has different interpretations for the peak and broadband (smooth) components. For the peak component, this corresponds to
\begin{equation}\label{alpha_p}
    \alpha_p(z)=\sqrt{\frac{D_M(z)D_H(z)/r_d^2}{[D_M(z)D_H(z)/r_d^2]_{\rm fid}}},
\end{equation}
where $r_d$ is the sound horizon scale at the end of the drag epoch. The broadband (or smooth) isotropic component $\alpha_s$ is less straightforward to model \citep{cuceu_cosmology_2021}, and is typically marginalized over as most of the cosmological information is contained within $\alpha_p$ and $\phi_f$, which contains the AP information from the full shape \citep{cuceu23b}. A measurement of the BAO scale corresponds to a constraint in the $\{H_0r_d,\Omega_\mathrm{m}\}$ parameter space. When including the full-shape information on $\Omega_\mathrm{m}$, this tightens constraints by a factor of $\sim2$ \citep{cuceu_cosmology_2021}. Since this additional constraining power comes from the broadband, the parameter $\phi_s$ is of particular interest. The key cosmological parameters in our compressed analysis are therefore $\phi_s$, $\phi_p$, and $\alpha_p$. We return to this in Section~\ref{sec:compressed}.

In order to proceed with the analysis, we need to model the power spectrum of the biased tracers. Following the modeling of \cite{bourboux_completed_2020} and \cite{cuceu_cosmology_2021}, also detailed in \cite{gerardi_2023}, the model for the power spectrum is
\begin{equation}\label{eqn:P_auto}
    P_{\rm Ly\alpha}(k,\mu_k,z)=b_\alpha^2 (1+\beta_\alpha\mu_k^2)^2 F_{\rm nl,Ly\alpha}^2(k,\mu_k)P(k,z),
\end{equation}
for the \lya forest and 
\begin{equation}\label{eqn:P_cross}
    \begin{split}
    P_\times(k,\mu_k,z) = b_\alpha(1+\beta_\alpha\mu_k^2)(b_Q+f(z)\mu_k^2) \times \\ F_{\rm nl, QSO}(k_\parallel)P(k,z)
    \end{split}
\end{equation}
for its cross-power spectrum with quasars. In each of the above equations, $P(k,z)$ is the linear isotropic matter power spectrum, and $\mu_k=k_\parallel/k$ where $k$ is the modulus of the wave vector and $k_\parallel$ is its projection along the line of sight. For \lya forest tracers, $b_\alpha$ is the linear bias of the \lya forest, and the \lya forest RSD parameter $\beta_\alpha=b_{\eta,\rm Ly\alpha} f(z)/b_\alpha$ is degenerate with an unknown velocity divergence bias $b_{\eta,\rm Ly\alpha}$ and the logarithmic growth rate $f(z)$ \citep{mcdonald_2003,slosar_2011,seljak_2012,givans_hirata_2020,chen_vlah_white_2021}. We treat both $b_\alpha$ and $\beta_\alpha$ as nuisance parameters to marginalize over. The $F_{\rm nl,Ly\alpha}$ term encodes nonlinear corrections modeled by \citep{arinyo_2015}, whose parameters we keep constant. We restrict our main analysis to linear scales ($r\geq30\,h^{-1 \rm}\,\rm Mpc$), so this choice should not significantly affect our results.

In the cross-power spectrum, the linear bias of quasars $b_Q$ is also relevant, and the RSD term for quasars is $\beta_Q=f(z)/b_Q$. We treat the QSO linear bias as a nuisance parameter and marginalize over it. Lastly, we model the effects of redshift errors and nonlinear peculiar velocities of quasars with the parameter $F_{\rm nl,QSO}(k_\parallel)$, which we define by the Lorentzian function used in \cite{bourboux_completed_2020},
\begin{equation}\label{eqn:fnl_qso}
    F_{\rm nl,QSO}(k_\parallel)=\sqrt{[1+(k_\parallel\sigma_z)^2]^{-1}},
\end{equation}
where the free parameter $\sigma_z$ captures the effects of redshift errors and the velocity dispersion. This is also a nuisance parameter that we marginalize over.

\subsection{Implementation}\label{sec:implementation}

We perform our analysis by fitting the models for the two-point correlation functions to the noiseless synthetic data defined by the fiducial cosmology with the appropriate covariance matrix. During the sampling process, \texttt{Vega}\footnote{\url{https://github.com/andreicuceu/vega}} \citep{cuceu_vega_2020} computes the theoretical correlation functions from the fiducial cosmology by decomposing the biased power spectrum into multipoles and transforming them into the correlation function with the FFTLog algorithm \citep{hamilton_2000}. \texttt{Vega} then jointly fits the model correlation functions to the noiseless synthetic data using the user-provided covariance matrix to estimate uncertainties on the nuisance parameters and either compressed (Section~\ref{sec:compressed}) or direct cosmological parameters (Section~\ref{sec:directfit}).

\subsubsection{Compressed Analysis}\label{sec:compressed}

In the compressed analysis, the correlation function is modeled by first decomposing the power spectrum into a peak (wiggles) and broadband (smooth) component \citep{kirkby_2013}. This is done because the BAO peak represents a well-defined feature in the power spectrum, and the decomposition allows us to determine how much additional information is coming from the broadband. Since the BAO peak is localized to $\sim100\,h^{-1}\rm Mpc$, the ability to constrain its position is robust against a variety of analysis choices and scale ranges. During the decomposition, nonlinear broadening of the BAO peak along and across the line of sight are also added as fixed parameters. The broadening is due to the nonlinear growth of structure, which is represented in the quasi-linear mock dataset \citep{casas_2025_validation}.

We use \texttt{Vega} to fit the model correlation functions to the noiseless synthetic data. We fit the compressed parameters $\{\alpha_p, \alpha_s, \phi_p, \phi_s\}$ where the $\alpha_{p,s}$ parameters denote the isotropic BAO scale (peak) or broadband (smooth) information, respectively, and $\phi_{p,s}$ denotes the same for the anisotropic (AP) information. We also fit and marginalize over the nuisance parameters $\{b_\alpha, \beta_\alpha, b_Q, \sigma_z, \Delta r_\parallel\}$. Table~\ref{tab:compressed} shows the fit parameters and their fiducial values, including the derived quantity $f\sigma_8$.

\begin{table}
\centering
\caption{The full set of parameters for the compressed analysis. The second column lists the fiducial values used to generate the synthetic correlations. The last four columns show the one-dimensional marginal constraints in different cases, where the scale indicated on the second line denotes either $r_\mathrm{max}=180$ or $240\,h^{-1}\,\mathrm{Mpc}$. The units for $\sigma_z$ and $\Delta r_\parallel$ are $h^{-1}\rm\,Mpc$.}
\label{tab:compressed}
\resizebox{\textwidth}{!}{
\begin{tabular}{lc|cccc}
\hline
Parameter & Fiducial & \multicolumn{4}{c}{68\% limits} \\
\cline{3-6}
& & \makecell{\footnotesize Distorted\\\scriptsize ($180\,h^{-1}\rm\,Mpc$)} &
\makecell{\footnotesize Undistorted\\\scriptsize ($180\,h^{-1}\rm\,Mpc$)} &
\makecell{\footnotesize Distorted\\\scriptsize ($240\,h^{-1}\rm\,Mpc$)} &
\makecell{\footnotesize Undistorted\\\scriptsize ($240\,h^{-1}\rm\,Mpc$)} \\
\hline
$\alpha_p$           & $1.0$     & $1.0\pm0.0061$ & $1.0\pm0.0059$ & $1.0\pm0.0060$ & $1.0\pm0.0057$ \\
$\alpha_s$           & $1.0$     & $1.0\pm0.012$ & $1.0\pm0.016$ & $1.0\pm0.012$ & $1.0\pm0.016$ \\
$\phi_p$             & $1.0$     & $1.0\pm0.021$ & $1.0\pm0.018$ & $1.0\pm0.020$ & $1.0\pm0.018$ \\
$\phi_s$             & $1.0$     & $1.0\pm0.0081$ & $1.0\pm0.0070$ & $1.0\pm0.0079$ & $1.0\pm0.0069$ \\
$f\sigma_8$          & $0.298$     & $0.298\pm0.025$ & $0.298\pm0.029$ & $0.298\pm0.024$ & $0.298\pm0.028$ \\
$b_\alpha$           & $-0.1352$ & $-0.1352\pm0.0014$ & $-0.1352\pm0.0041$ & $-0.1352\pm0.0014$ & $-0.1352\pm0.0040$ \\
$\beta_\alpha$       & $1.445$   & $1.445\pm0.027$ & $1.445\pm0.041$ & $1.445\pm0.025$ & $1.445\pm0.040$ \\
$b_{\rm Q}$          & $3.545$   & $3.545\pm0.041$ & $3.545\pm0.10$ & $3.545\pm0.038$ & $3.545\pm0.10$ \\
$\sigma_z$           & $3.18$    & $3.18\pm0.82$ & $3.18\pm0.92$ & $3.18\pm0.79$ & $3.18\pm0.90$ \\
$\Delta r_\parallel$ & $0.0$     & $0.0\pm0.13$ & $0.0\pm0.14$ & $0.0\pm0.13$ & $0.0\pm0.13$ \\
\hline
\end{tabular}
}
\end{table}

\subsubsection{Direct Fit}\label{sec:directfit}

In the direct fit framework, we sample cosmological parameters through a direct fit of the full shape of the 3D correlation functions. We use a custom likelihood for the \texttt{Cobaya} framework \citep{torrado_lewis_cobaya_2019, torrado_lewis_cobaya_2021} which uses \texttt{CAMB} \citep{lewis_camb_2000} to compute the linear matter power spectrum and interfaces with \texttt{Vega} to model the correlation functions. We perform a joint fit to the full shape of the \lya autocorrelation and \lya $\times$ QSO cross-correlation by sampling the parameter space with Markov chain Monte Carlo \citep[MCMC;][]{lewis_bridle_mcmc_2002, neal_mcmc_2005, lewis_mcmc_2013}.

At each sampling step, Cobaya generates a template power spectrum over the range $10^{-3}<k<10 ~h ~\rm Mpc^{-1}$. \texttt{Vega} transforms this full power spectrum into correlation function models for the \lya auto- and cross-correlation without any separation into peak and smooth components. Because the decomposition is skipped in the direct fit analysis, this means that we do not add any nonlinear broadening to the BAO peak. The cosmological parameters we sample are $\{H_0, \Omega_\mathrm{m}, \Omega_\mathrm{b}h^2, A_s, n_s\}$, where $H_0$ is the Hubble constant, $\Omega_\mathrm{m}$ is the present-day matter density parameter, $\Omega_\mathrm{b}$ is the present-day baryon density parameter, and $A_s$ and $n_s$ are the amplitude and slope of the primordial power spectrum, respectively. We also sample the nuisance parameters $\{b_\alpha, \beta_\alpha, b_Q,\sigma_v\}$ and marginalize over them. Table~\ref{tab:directfit} lists the parameters we sample with their fiducial values and priors. Any parameters that we do not explicitly sample are fixed to their Planck 2018 values \citep{planck_2018_cosmology}.

\begin{table}
	\centering
	\caption{The full set of sampled parameters for direct cosmological inference. The second column lists the fiducial values used to generate the synthetic correlations. The priors for sampling are shown in the third column, where $\mathcal{U}$ denotes a uniform prior and the `$\log$' subscript indicates sampling in logarithmic space. We provide the one-dimensional marginal constraints in the last four columns for different cases, where the value corresponds to the mean of the posterior. The scale indicated on the second line denotes either $r_\mathrm{max}=180$ or $240\,h^{-1}\,\mathrm{Mpc}$. The units for $H_0$ are $\rm km/s/Mpc$ and the units for $\sigma_z$ are $h^{-1}\rm\,Mpc$.}
	\label{tab:directfit}
    \resizebox{\textwidth}{!}{
	\begin{tabular}{lcc|cccc}
		\hline
		Parameter & Fiducial & Prior & \multicolumn{4}{c}{68\% limits} \\
		\cline{4-7}
		         &          &       & \makecell{\footnotesize Distorted\\\scriptsize ($180\,h^{-1}\rm\,Mpc$)} & \makecell{\footnotesize Undistorted\\\scriptsize ($180\,h^{-1}\rm\,Mpc$)} & \makecell{\footnotesize Distorted\\\scriptsize ($240\,h^{-1}\rm\,Mpc$)} & \makecell{\footnotesize Undistorted\\\scriptsize ($240\,h^{-1}\rm\,Mpc$)} \\
		\hline
        $H_0$   & $67.36$     & $\mathcal{U}(40, 100)$   & $69.7^{+2.1}_{-5.4}$ & $69.8^{+2.3}_{-5.6}$ & $69.5^{+2.0}_{-5.0}$ & $69.7^{+2.3}_{-5.2}$ \\
        $\Omega_\mathrm{m}$   & $0.31509$   & $\mathcal{U}(0.01, 0.99)$  & $0.318\pm 0.011$ & $0.318\pm 0.010$ & $0.318\pm 0.011$ & $0.318\pm 0.010$ \\
        $\Omega_\mathrm{b}h^2$              & $0.02237$   & $\mathcal{U}(0.01, 0.05)$            & $0.0253^{+0.0022}_{-0.0064}$ & $0.0255^{+0.0024}_{-0.0066}$ & $0.0251^{+0.0020}_{-0.0060}$ & $0.0253^{+0.0024}_{-0.0062}$ \\
        $10^9 A_s$                      & $2.1$ & $\mathcal{U}_{\log(10^{10}A_s)}(0.5, 6)$ & $1.98^{+0.29}_{-0.37}$ & $1.97^{+0.32}_{-0.45}$ & $1.99^{+0.29}_{-0.35}$ & $1.97^{+0.32}_{-0.44}$ \\
        $n_s$                      & $0.9649$    & $\mathcal{U}(0.8, 1.2)$              & $0.950^{+0.035}_{-0.020}$ & $0.946^{+0.042}_{-0.030}$ & $0.951^{+0.032}_{-0.019}$ & $0.947^{+0.040}_{-0.029}$ \\
		$b_\alpha$                & $-0.1352$   & $\mathcal{U}_{\log(-b_\alpha)}(-2, 0)$& $-0.138^{+0.012}_{-0.0099}$ & $-0.138^{+0.014}_{-0.011}$ & $-0.137^{+0.012}_{-0.0094}$ & $-0.138^{+0.014}_{-0.011}$ \\
        $\beta_\alpha$             & $1.445$     & $\mathcal{U}(0, 5)$                  & $1.448\pm 0.024$ & $1.458\pm 0.046$ & $1.448\pm 0.022$ & $1.458\pm 0.045$ \\
        $b_{Q}$                & $3.545$     & $\mathcal{U}_{\log(b_Q)}(-2, 1.3)$              & $3.62^{+0.28}_{-0.35}$ & $3.64^{+0.35}_{-0.46}$ & $3.61^{+0.26}_{-0.33}$ & $3.64^{+0.35}_{-0.45}$ \\
        $\sigma_z$ & $3.18$      & $\mathcal{U}(0, 15)$                 & $2.84^{+1.1}_{-0.67}$ & $2.77^{+1.2}_{-0.78}$ & $2.87^{+1.1}_{-0.65}$ & $2.79^{+1.2}_{-0.75}$ \\
		\hline
	\end{tabular}
    }
\end{table}

\section{Compressed Analysis Results}\label{sec:results_compressed}

In this section, we present the results of our cosmological forecasts on compressed parameters using \texttt{Vega}. The following subsections explore three main categories of changes to the standard analysis: the impact of using the true continuum (Section~\ref{sec:truecont}), variations in the scale range (Section~\ref{sec:scalecuts}), and modifications to the covariance estimation (Section~\ref{sec:cov_mods}).

Throughout this section, we fit the isotropic scale parameters $\{\alpha_p, \alpha_s\}$ and the AP scale parameters $\{\phi_p, \phi_s\}$. We marginalize over the nuisance parameters.

\subsection{True Continuum}\label{sec:truecont}

The true continuum should provide better cosmological measurements, at least on scales that are not measured well in the continuum fitting approach. The exact scales impacted by the continuum fitting are difficult to quantify, as the effect is not localized in real space. If all forests were the same length, the continuum fitting procedure would erase all power on the scale of the forest length and larger, resulting in a sharp cutoff at this low-$k$ mode in the power spectrum. For example, in the mock datasets employed here, the maximum forest length is $\sim450\,h^{-1}\,\mathrm{Mpc}$, corresponding to a complete loss of power at $k\lesssim0.014\,h\,\mathrm{Mpc}^{-1}$ in the distorted analysis. In practice, all forests are different lengths, resulting in a more gradual effect over a range of scales \cite{blomqvist_2015}.

\begin{figure*}
    \centering
    \includegraphics[width=\textwidth]{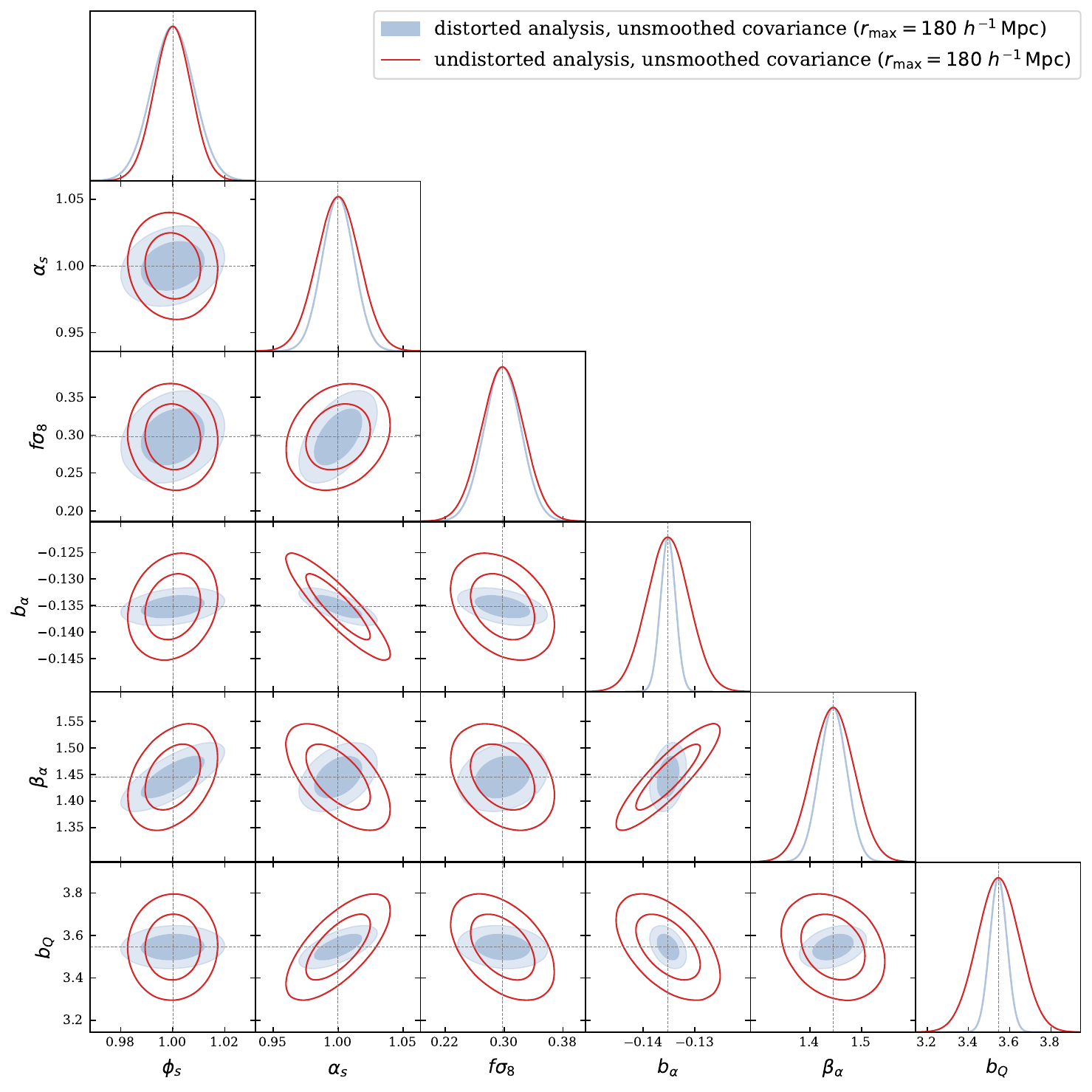}
    \caption{Forecast constraints on compressed cosmological parameters using two analyses at the standard scale range ($r_\mathrm{max}=180\,h^{-1}\,\mathrm{Mpc}$) with unsmoothed covariances: the continuum fitting (distorted) analysis (\textit{solid blue contours}) and the true continuum (undistorted) analysis (\textit{red contours}). This comparison isolates the impact of continuum fitting versus knowledge of the true continuum, as both analyses use identically computed covariance matrices with the same corrections applied. The advantage of the true continuum is evident in the broadband AP scale parameter, $\phi_s$. The BAO AP scale parameter, $\phi_p$, also improves with the true continuum but is not shown here (see Table~\ref{tab:compressed}). See Sections~\ref{sec:forestcontinuum} and \ref{sec:truecont} for further discussion.}
    \label{fig:compressed_truecont}
\end{figure*}

Figure~\ref{fig:compressed_truecont} shows the forecast results from the distorted and undistorted analyses with unsmoothed covariances computed in the same way. Since these forecasts also employ the same scale cuts, this isolates the effect of the true continuum. We show constraints on three cosmological parameters ($\phi_s,\alpha_s,f\sigma_8$) and three nuisance parameters ($b_\alpha,\beta_\alpha,b_Q$). The greatest improvement is seen in the broadband AP scale parameter $\phi_s$, where the true continuum enables a nearly $15\%$ improvement in constraining power over continuum fitting. The constraint on the BAO AP parameter $\phi_p$ also improves by $\sim10\%$ with the true continuum, though this is not shown here.

Figure~\ref{fig:compressed_truecont} also shows constraints on $f\sigma_8$, the linear biases, and the RSD parameter, all of which degrade when using the true continuum. In several cases, the degeneracy directions shift between the distorted and undistorted analyses, altering how information is recovered. The distortion matrix and projection formalism rely on a number of assumptions that have been validated for the BAO and AP parameters, but not yet for $\alpha_s$ or the bias terms \cite{cuceu23b,cuceu_desi_lya_validation,cuceu_2025}. Additionally, the distorted analysis allows small-scale information below $r_\mathrm{min}$ to leak into the fit \cite{busca_distortion_matrix}. This could cause a significant improvement in constraining power in the context of a forecast which fits an idealized synthetic data vector built from the model. In practice, we cannot yet model these small scales sufficiently well. As a result, constraints on these parameters may be biased in the distorted analysis, which could translate to a bias in the inferred growth rate, as seen in \cite{cuceu_2025}. This also implies that our forecast improvement on AP parameters with the true continuum could be underestimated. Importantly, the \lya forest is not a competitive probe of $f\sigma_8$ compared to galaxy surveys, and its most impactful cosmological constraint beyond BAO comes from the AP effect. Table~\ref{tab:compressed} summarizes the one-dimensional marginal constraints on all parameters for two scale ranges in each analysis type. 

The improvement on $\phi_s$ is also shown in Figure~\ref{fig:rmin_rmax_improvement}, where the undistorted analysis yields the tightest constraints across all tested scale cuts. As previously mentioned, the true continuum analysis achieves $\sim15\%$ tighter constraints than the unsmoothed distorted analysis, and this relative improvement remains roughly constant even as $r_\mathrm{max}$ increases. This consistency may be in part due to the fact that most \lya forests in our mock dataset have lengths exceeding the largest scales probed here. Relative to the current baseline with a smoothed covariance, the improvement from the true continuum is $\sim10\%$. This is in addition to the improvement with extensions to larger scales (see Section~\ref{sec:scalecuts}).

The $\sim15\%$ improvement on $\phi_s$ corresponds to a covariance rescaling factor of $f_{\rm cov}\approx0.85^2$ in the below relation (adapted from Eqn. 15 in \cite{cuceu_cosmology_2021}):
\begin{equation}\label{fcov}
    f_{\rm cov} \propto \frac{A_{\rm distorted}}{A_{\rm undistorted}},
\end{equation}
where $A_\mathrm{undistorted}$ and $A_\mathrm{distorted}$ refer to the survey area in the undistorted and distorted analysis, respectively. Therefore, a $15\%$ reduction in uncertainty corresponds to the constraining power achievable with an additional $\sim40\%$ of \lya forest survey area.

\subsection{Changes in Scale Range}\label{sec:scalecuts}

\begin{figure}
    \centering
    \includegraphics[width=\textwidth]{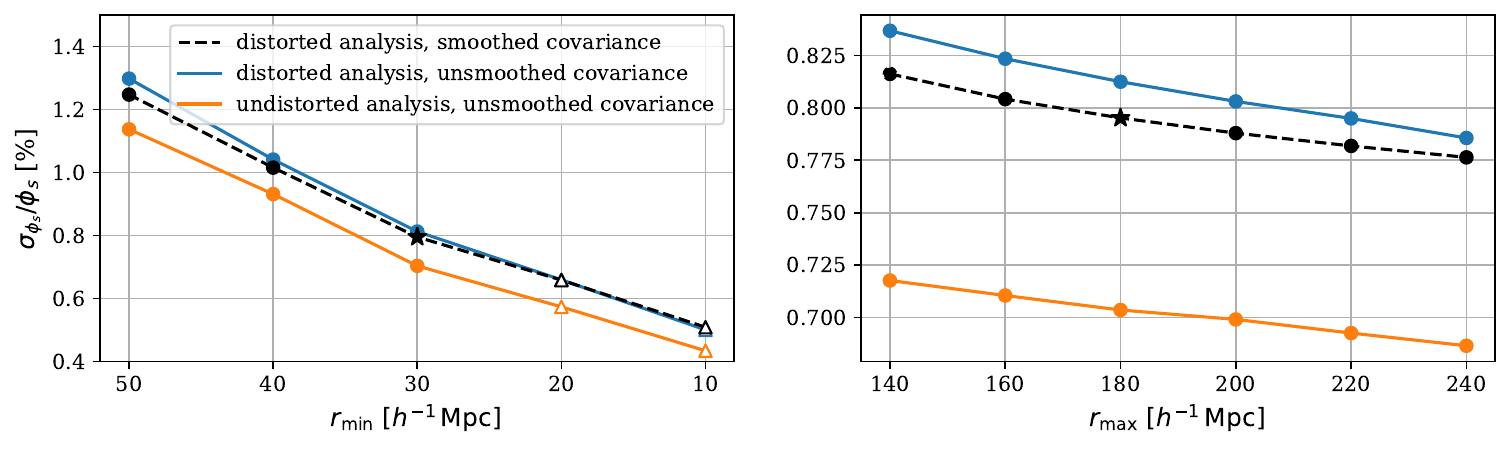}
    \caption{Forecast $1\sigma$ precision (\%) on the AP broadband scale parameter $\phi_s$ for changes in the minimum and maximum scale range of the correlation function. The left-hand panel shows these as a function of the minimum isotropic scale cut $r_\mathrm{min}$: filled circles indicate a common cut for both the auto- and cross-correlation, while open triangles denote the specified cut in the auto-correlation and a fixed $r_\mathrm{min} = 30\,h^{-1}\,\mathrm{Mpc}$ in the cross-correlation. The right-hand panel shows the precision as a function of the maximum isotropic scale cut $r_\mathrm{max}$. Results from three analysis types are shown: the distorted analysis with a smoothed covariance matrix (\textit{black dashed line}), the distorted continuum fitting analysis with an unsmoothed covariance matrix (\textit{blue line}), and the undistorted true continuum analysis with an unsmoothed covariance matrix (\textit{orange line}). The latter two analyses use covariances computed in the same way and include the Hartlap correction in the likelihood analysis. The current baseline analysis is indicated by the black stars and corresponds to the distorted analysis with a smoothed covariance and scale cuts of $r_\mathrm{min} = 30\,h^{-1}\,\mathrm{Mpc}$ and $r_\mathrm{max} = 180\,h^{-1}\,\mathrm{Mpc}$.}
    \label{fig:rmin_rmax_improvement}
\end{figure}

In addition to changes in the analysis method, additional cosmological information should be accessible when extending the analysis to a broader range of scales. To quantify the information gain from different physical scales, we examine how forecast constraints on $\phi_s$ change when varying the minimum and maximum separations used in the analysis. The current baseline continuum fitting analysis typically adopts a scale range of $r=30-180\,h^{-1}\,\mathrm{Mpc}$ \citep{desi_dr2_lya}. Figure~\ref{fig:rmin_rmax_improvement} shows the improvements in constraints on $\phi_s$ as a function of these isotropic scale cuts, comparing the current baseline analysis, the distorted analysis using an unsmoothed covariance, and the true continuum (undistorted) analysis with an unsmoothed covariance. 

In the left-hand panel of Figure~\ref{fig:rmin_rmax_improvement}, we vary $r_\mathrm{min}$ while fixing the maximum isotropic separation to $r_\mathrm{max}=180\,h^{-1}\,\mathrm{Mpc}$. The open triangles for $r_\mathrm{min}=10 ~\mathrm{and}~ 20 \,h^{-1}\,\mathrm{Mpc}$ indicate that the $r_\mathrm{min}$ choice was adopted for the auto-correlation but fixed at $r_\mathrm{min}=30\,h^{-1}\,\mathrm{Mpc}$ for the cross-correlation. This is motivated by the fact that nonlinearities are more difficult to accurately model in the cross-correlation due to the stronger bias of quasars relative to the forest, whereas modeling these smaller scales may be more feasible for the auto-correlation. As shown in the figure, decreasing $r_\mathrm{min}$ results in similar improvements for all three analyses. Specifically, reducing $r_\mathrm{min}$ from $50$ to $10\,h^{-1}\,\mathrm{Mpc}$ (in the auto-correlation) leads to a $\sim60\%$ improvement on $\phi_s$. Relative to the typical baseline analysis choice of $r_\mathrm{min}=30\,h^{-1}\,\mathrm{Mpc}$, the improvement enabled by decreasing the minimum scale cut to $r_\mathrm{min}=10\,h^{-1}\,\mathrm{Mpc}$ in the auto-correlation is $\sim40\%$, or $\sim50\%$ when also exploiting the true continuum. However, for consistency, we adopt the standard minimum scale cut of $r_\mathrm{min}=30\,h^{-1}\,\mathrm{Mpc}$ for the main results in this paper.

The right-hand panel of Figure~\ref{fig:rmin_rmax_improvement} shows the effect of increasing the maximum isotropic separation $r_\mathrm{max}$ while fixing the minimum scale cut to $r_\mathrm{min}=30\,h^{-1}\,\mathrm{Mpc}$. Increasing $r_\mathrm{max}$ from $140$ to $240\,h^{-1}\,\mathrm{Mpc}$ results in a $\sim5\%$ gain in information on $\phi_s$, including for the distorted analysis. The advantage of the undistorted analysis over the distorted case remains roughly constant as $r_\mathrm{max}$ increases, as described in Section~\ref{sec:truecont}. For the main results presented in this paper, we adopt maximum scale cuts of $r_\mathrm{max}=180$ and $240\,h^{-1}\,\mathrm{Mpc}$.

\begin{figure}
    \centering
    \includegraphics[width=\linewidth]{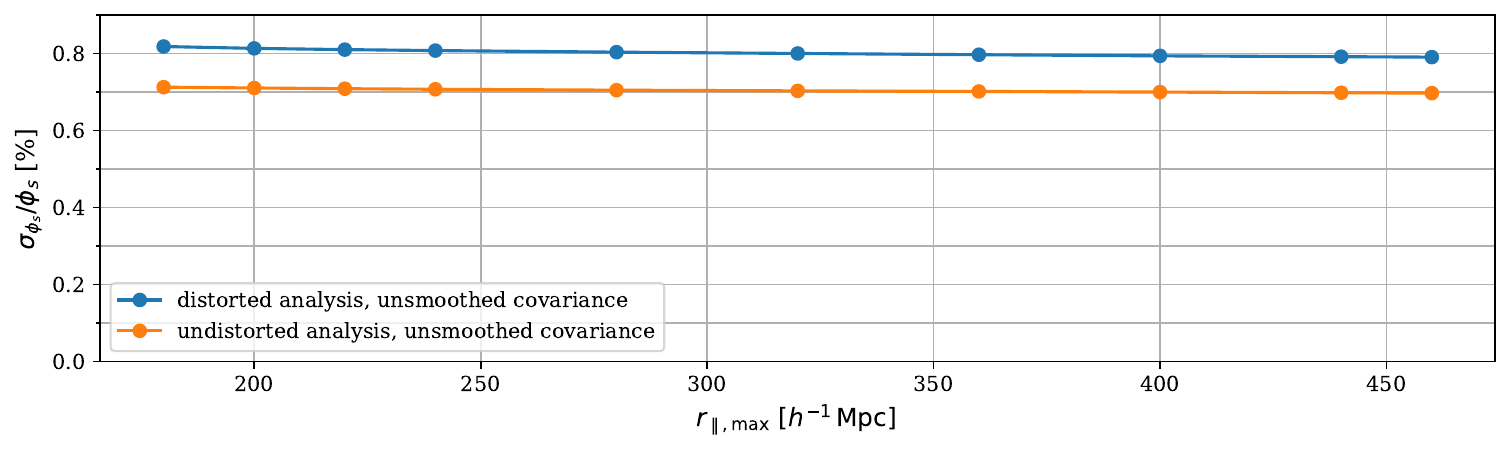}
    \caption{Forecast $1\sigma$ precision (\%) on the AP broadband parameter $\phi_s$ with different line-of-sight scale cuts $r_{\parallel,\mathrm{max}}$. The point at $r_\mathrm{max}=180\,h^{-1}\,\mathrm{Mpc}$ uses this cut both along and across the line of sight, while the remaining points fix the maximum transverse separation to $r_{\perp,\mathrm{max}}=200\,h^{-1}\,\mathrm{Mpc}$. The distorted (\textit{blue}) and undistorted (\textit{orange}) analyses both use unsmoothed covariances computed up to separations of $r_{\parallel,\mathrm{max}}=460\,h^{-1}\mathrm{Mpc}$ along and $r_{\perp,\mathrm{max}}=200\,h^{-1}\mathrm{Mpc}$ across the line of sight. This is in contrast to Figure~\ref{fig:rmin_rmax_improvement}, where larger separations are probed in both directions. The undistorted analysis consistently outperforms the distorted case at all scale cuts tested, but information gain at very large scales is minimal.}
    \label{fig:AP_large_scales}
\end{figure}

We also investigate whether additional information can be extracted at even larger scales. To do this, we compute covariance matrices from correlation functions measured up to $r_{\parallel,\mathrm{max}} = 460\,h^{-1}\,\mathrm{Mpc}$ along the line of sight and $r_{\perp,\mathrm{max}} = 200\,h^{-1}\,\mathrm{Mpc}$ in the transverse direction. These anisotropic limits are chosen to ensure a sufficient number of independent sky regions when using $\texttt{NSIDE}=16$ HEALPix pixels. In this case, the unsmoothed covariance is estimated from a stack of 150 mocks to compensate for the larger size of the data vector. The results of this test, for both the distorted and undistorted cases, are shown in Figure~\ref{fig:AP_large_scales}. We find that the undistorted analysis consistently outperforms the distorted case by $\gtrsim10\%$ across all tested scale cuts. The shallow slope of each curve indicates that only minimal additional information is gained at the largest scales; however, separations beyond $200\,h^{-1}\,\mathrm{Mpc}$ are only probed in the line-of-sight direction in this test.

\subsection{Covariance Modifications}\label{sec:cov_mods}

We also explore the effects of choices we make to estimate the covariance. One key choice for estimating the covariance is the size of the HEALPix pixels used to compute individual correlation functions, since this affects the number of samples $N$, where $N=N_\mathrm{mocks}\times N_\mathrm{HEALPix}$. In our main analysis, we fix the number of mocks and the HEALPix size as described in Section~\ref{sec:covariance}. The other choice directly affecting the covariance matrix is whether or not to apply the smoothing algorithm that is currently used in the baseline analysis.

In Figure~\ref{fig:rmin_rmax_improvement}, the smoothed (current baseline) and unsmoothed distorted analyses show similar constraints ($\lesssim3\%$) on the AP scale parameter. Therefore, the impact of covariance smoothing in the baseline distorted analysis is insignificant. Although not shown, we did not find a significant difference in the performance on the isotropic BAO parameter $\alpha_p$ between the current baseline and the unsmoothed distorted analysis.

\section{Direct Cosmology Results}\label{sec:results_direct}

\begin{figure*}
    \centering
    \includegraphics[width=\textwidth]{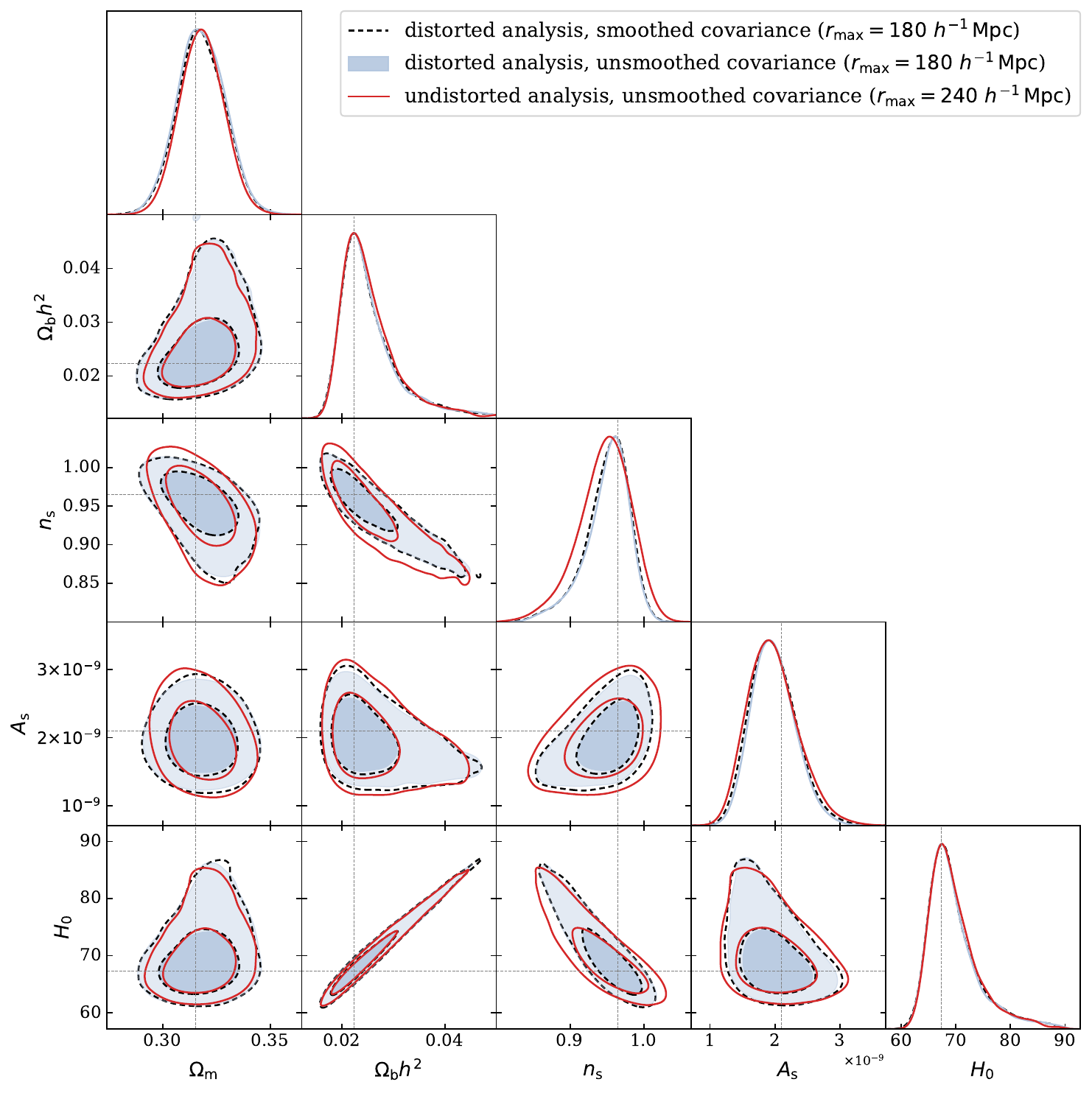}
    \caption{Forecast constraints on direct cosmological parameters obtained using MCMC sampling with \texttt{Cobaya}. We compare results from the continuum fitting (distorted) analysis with an unsmoothed (\textit{solid blue contours}) or smoothed (\textit{dashed black contours}) covariance and the standard maximum isotropic scale cut of $r_\mathrm{max}=180\,h^{-1}\,\mathrm{Mpc}$ to those from the undistorted, unsmoothed covariance analysis with an extended scale range of $r_\mathrm{max}=240\,h^{-1}\,\mathrm{Mpc}$ (\textit{red contours}). The latter represents the best possible improvement on recovered cosmological information studied in this paper. We find the largest improvement on $\Omega_\mathrm{m}$ (approximately $10\%$), which is consistent with our findings for compressed parameters in Section~\ref{sec:results_compressed}. The degradation of constraints on other parameters with the true continuum is likely a consequence of the sub-optimal correlation function estimator (see Sections~\ref{sec:covariance} and \ref{sec:discussion}).}
    \label{fig:direct_cosmo}
\end{figure*}

While the compressed parameters provide a valuable comparison point to models, they may not capture all of the cosmological information \citep{gerardi_2023}. We therefore also perform direct cosmological inference with \texttt{Cobaya}. Figure~\ref{fig:direct_cosmo} shows our forecasts for cosmological parameters from a direct fit to the full shape of the correlation functions. This figure shows results from the standard scale range of $r=30-180\,h^{-1}\,\mathrm{Mpc}$ for the distorted analysis with smoothed and unsmoothed covariances, and a larger scale range of $r=30-240\,h^{-1}\,\mathrm{Mpc}$ for the undistorted analysis. Table~\ref{tab:directfit} lists the one-dimensional marginal constraints for the two different scale ranges for each of the unsmoothed analyses.

We find a $\sim10\%$ improvement in $\Omega_\mathrm{m}$ with the undistorted analysis, consistent with our results from the compressed forecasts. Interestingly, the constraints on the other flat $\Lambda$CDM cosmological parameters worsen in the undistorted case. This may be related to the degradation of bias parameter constraints when using the true continuum, as described in Section~\ref{sec:truecont}. However, the \lya forest alone provides only modest constraining power on these remaining parameters, which are much more tightly constrained in combination with other probes such as the cosmic microwave background \citep[e.g.][]{gerardi_2023}.

\section{Discussion}\label{sec:discussion}

We showed that knowledge of the true continuum leads to approximately $10\%$ tighter constraints on the AP scale parameter and $\Omega_\mathrm{m}$ compared to the current continuum fitting approach in the idealized case without contaminants. This improvement holds across all scale cuts tested. Moreover, additional constraining power is accessible at larger scales in both the continuum fitting (distorted) and true continuum (undistorted) analyses, leading to an overall $\sim15\%$ improvement relative to the current baseline when extending the true continuum analysis to $r_\mathrm{max}=240\,h^{-1}\,\mathrm{Mpc}$. Interestingly, forecasts on other parameters such as $f\sigma_8$, $n_s$, and $A_s$ revealed a degradation in constraining power with the true continuum; however, these parameters are not well constrained by the \lya forest and may be prone to bias in the distorted analysis.

The results presented in this work rely on accurate estimates of the covariance matrix. As discussed in Section~\ref{sec:covariance}, the current estimators for the \lya two-point correlation function and its covariance are sub-optimal. While they have been extensively validated in the baseline distorted analysis, their impact in an undistorted analysis must be studied further. It is possible that the projection inherent in the distorted analysis results in artificially tight constraints on nuisance parameters and $f\sigma_8$, for example because the distorted analysis allows small-scale information to leak into the fit \cite{busca_distortion_matrix}. As a result, our measured improvements in the AP parameters with the true continuum may be conservative. Directly testing the impact of the sub-optimal estimator would require the development and validation of an optimal estimator. If one wishes to constrain RSD from the 3D \lya forest, our results suggest that tighter constraints than the current baseline will first require the development of an optimal estimator for these parameters. In contrast, we find that access to the true continuum could already enable a $\sim$10–15\% improvement on the AP effect with the current estimators.

As shown in Figure~\ref{fig:compressed_truecont}, the undistorted analysis yields different degeneracy directions between several parameters. For instance, the correlation coefficient between the linear bias $b_\alpha$ and the RSD parameter $\beta_\alpha$ shifts from roughly $0.29$ in the distorted analysis to $0.88$ in the undistorted case. Perhaps more importantly, the correlation coefficient between $\phi_s$ and $\alpha_s$ reduces from approximately $0.22$ to $-0.08$. This may be particularly beneficial for a full-shape cosmological analysis on data, as $\alpha_s$ may be biased by large-scale distortions introduced in the continuum fitting analysis. These changes in degeneracy structure may also be related to the correlation function estimator and the projection in the distorted analysis. Different degeneracy directions between nuisance parameters can also influence the amount of cosmological information that may be present on different scales. Additionally, the cosmological information appears to be highly sensitive to the off-diagonal structure of the covariance matrix. For example, ignoring the cross-covariance between the auto- and cross-correlations in both the distorted baseline and undistorted analyses reduces the relative improvement on $\phi_s$ enabled by the true continuum from approximately $14\%$ to $8\%$. This impact is lessened when comparing instead to the unsmoothed distorted case, where the relative improvement drops more modestly from roughly $16\%$ to $12\%$. These shifts highlight the importance of including the cross-covariance to obtain more accurate uncertainty estimates and fully realize the benefits of the true continuum.

There may be even more cosmological information accessible at larger scales. We tested this with forecasts using covariance matrices computed up to line-of-sight separations of $r_{\parallel,\mathrm{max}}=460\,h^{-1}\,\mathrm{Mpc}$, while we fixed the maximum transverse separation at $r_{\perp,\mathrm{max}}=200\,h^{-1}\,\mathrm{Mpc}$ to avoid even larger HEALPix pixels and thus fewer subsamples for the covariance matrix measurement. While we found hints of extra information up to $r_\mathrm{max}\approx450\,h^{-1}\,\mathrm{Mpc}$, the relative gain is modest, likely due to the weak clustering signal and the fact that we are only probing these large scales along the line-of-sight. 

There are several ongoing and upcoming spectroscopic surveys such as DESI (including DESI-II \citep{schlegel_desi2_2025}), Euclid \citep{euclid_overview_2025}, and Roman \citep{roman_highlat_2022} that will extend our reach into the $z>2$ universe. However, the next major advance in \lya forest datasets is expected to come from a Stage-V spectroscopic survey with Spec-S5 \citep{spec-s5_2025}, anticipated in about a decade. In this context, developing new analysis methods that extract additional information from existing and forthcoming data can significantly enhance the scientific return of these surveys. This is particularly timely given the emerging tensions with $\Lambda$CDM \citep{desi_y1_7,desi_dr2_discretetracers,des_bao_sn_2025}.

Full-shape measurements of the \lya correlation functions using the new avenues explored in this work offer a promising path to extract more information from current datasets. To realize the gains suggested by this work, several future directions are essential. These include validating that predictions of the true continuum yield accurate two-point correlation measurements, and improving the estimation of the covariance matrix especially with extensions to larger scales. This may entail alternative binning strategies and validation of mock-based covariance estimates. Since we only observe one realization of the universe, it may be necessary to compute the covariance matrix with mocks or analytically, rather than relying on the assumptions currently employed to smooth the data covariance matrix (see Section~\ref{sec:covariance}).

It may also be worthwhile to explore full-shape analyses in multipoles, as this would greatly simplify the covariance estimation and make the gains enabled by the true continuum more tractable. In addition, the true continuum could potentially enhance measurements of the three-dimensional power spectrum, especially in light of recent developments in optimal estimators \citep[e.g.][]{roger_p3d_2024,karacayli_p3d_2025}. Furthermore, extending the analysis to larger scales ($r_\mathrm{max}>180\,h^{-1}\,\mathrm{Mpc}$) will require careful assessment and accurate modeling of potential systematics at these scales, such as UV background fluctuations \citep[e.g.][]{croft_2004,pontzen_2014a,gontcho_a_gontcho_2014,pontzen_2014b} and He\,\textsc{ii} reionization \citep[e.g.][]{furlanetto_2008,becker_2011,worseck_2016}. It will also be important to develop a more optimal estimator of the two-point correlation function, and ideally include the effects of continuum errors in this estimator. This would enable a more robust study of how continuum prediction errors impact cosmological constraints. In the context of the standard continuum fitting approach, the distortion matrix relies on several assumptions that have so far only been tested within the regime typically studied \citep{busca_distortion_matrix}. Extending the analysis to new regimes will require validating the distortion matrix in those contexts, and may necessitate modifications or more complex algorithms if the underlying assumptions no longer hold. For example, it is necessary to compute the distortion matrix up to sufficiently large line-of-sight separations ($r_\parallel=300\,h^{-1}\,\mathrm{Mpc}$) to accurately model the auto-correlation up to $r_\parallel=200\,h^{-1}\,\mathrm{Mpc}$, but this is still insufficient for modeling the cross-correlation to $r_\parallel=200\,h^{-1}\,\mathrm{Mpc}$ \cite{busca_distortion_matrix}. The reliability of the distortion matrix at large scales will be an important consideration for extending the analysis to $r>200\,h^{-1}\,\mathrm{Mpc}$. However, our results indicate that most of the improvement in AP arises from the use of the true continuum, making this approach particularly promising.

Our work focused exclusively on the idealized case of \lya forest measurements without contamination from metal absorption lines, optically thick absorbers, or broad absorption lines in the forest, and assuming perfect knowledge of the true continuum. Future studies will need to investigate the impact of these contaminants on the potential gains from the true continuum. Additionally, in practice, any continuum prediction method will introduce errors that will degrade our results. For example, current LyCAN predictions exhibit continuum errors at the $\sim2\%$ level \citep{turner_2024}. Quantifying the impact of these errors on cosmological constraints will be an important direction for future work, as the exact level of degradation will depend on whether the errors are correlated and how well we can model them. We plan to use improved versions of LyCAN to measure the two-point correlation functions with DESI data in a future paper, enabling the next step toward full-shape \lya forest cosmology with the true continuum.

\section{Summary}\label{sec:summary}

We presented the first forecasts with the \lya forest auto-correlation and its cross-correlation with quasars that are unaffected by continuum fitting distortion. We performed forecasts on compressed cosmological parameters through maximum likelihood estimation with \texttt{Vega} and sampled posteriors of direct cosmological parameters using a custom likelihood for \texttt{Cobaya} that interfaces with \texttt{Vega}. Given a model power spectrum, \texttt{Vega} creates a model for the two-point correlation functions and fits the model to the noiseless synthetic data defined by our fiducial cosmology.

To forecast constraints on cosmological parameters, we required estimates of the covariance matrix for the distorted and undistorted analyses. To study the additional constraining power from larger scales, we computed covariance matrices over separations from $r=30-240\,h^{-1}\rm\,Mpc$ using 50 mocks with HEALPix pixels covering areas of linear size $\sim280\,h^{-1}\rm\,Mpc$. We applied the Hartlap \citep{hartlap_2007} correction to the inverse covariance matrix in our likelihood analysis and validated its robustness for our specific regime of sample size and number of data bins in Appendix~\ref{sec:appendix_hartlap}.

We found that using the true continuum and extending the analysis to $r_\mathrm{max}=240\,h^{-1}\,\mathrm{Mpc}$ can enable up to $\sim15\%$ tighter constraints on the AP scale parameter and $\Omega_\mathrm{m}$ when compared to the baseline continuum fitting analysis at $r_\mathrm{max}=180\,h^{-1}\,\mathrm{Mpc}$. This improvement is analogous to an extra $\sim40\%$ of \lya forest survey area. When not exploiting larger scales and instead comparing constraints using the same standard scale range of $r=30-180\,h^{-1}\rm\,Mpc$, knowledge of the true continuum still results in $\sim10\%$ tighter constraints on these parameters, analogous to an additional $\sim25\%$ of \lya forest survey area.

To explore even larger scales, we extended our forecasts on compressed parameters to include separations up to $r_\mathrm{max} = 460\,h^{-1}\,\mathrm{Mpc}$. This required computing covariance matrices from a larger set of mocks (150), using HEALPix pixels with areas of linear size $\sim250\,h^{-1}\,\mathrm{Mpc}$. This setup allowed us to probe scales up to $r_{\parallel,\mathrm{max}}=460\,h^{-1}\,\mathrm{Mpc}$ along the line of sight, while limiting the transverse separations to $r_{\perp,\mathrm{max}}=200\,h^{-1}\,\mathrm{Mpc}$. We found evidence for additional constraining power at these larger scales; however, extending only the line-of-sight range provided minimal improvement.

This work has significant implications for cosmology. Recent DESI results have found evidence for significant tension with the Planck $\Lambda$CDM model \citep{desi_y1_7,desi_dr2_discretetracers}. While results from the full five-year DESI survey (DR3) may increase the statistical significance of this tension, future major advancements in cosmology will arise from next-generation surveys including Roman, Rubin, and Spec-S5. In the meantime, meaningful progress will rely on improved analysis techniques that can extract more information from existing data. The forecast improvement in the Alcock-Paczyński parameter from this work suggests that the \lya forest may soon have the ability to help distinguish between the best-fit $\Lambda$CDM and $w_0w_a$CDM models. Further, the improved precision on $\Omega_\mathrm{m}$ offers a promising avenue for shedding light on the emerging tension between DESI and Planck measurements within the $\Lambda$CDM model.

\section*{Data Availability}
Data points from each figure are available at \href{https://doi.org/10.5281/zenodo.19927452}{https://doi.org/10.5281/zenodo.19927452}.

\acknowledgments
We thank Naim Karaçaylı and David Weinberg for helpful discussions. WT and PM acknowledge support from the United States Department of Energy, Office of High Energy Physics under Award Number DE-SC-0011726. AC acknowledges support provided by NASA through the NASA Hubble Fellowship grant HST-HF2-51526.001-A awarded by the Space Telescope Science Institute, which is operated by the Association of Universities for Research in Astronomy, Incorporated, under NASA contract NAS5-26555. 

This material is based upon work supported by the U.S. Department of Energy (DOE), Office of Science, Office of High-Energy Physics, under Contract No. DE–AC02–05CH11231, and by the National Energy Research Scientific Computing Center, a DOE Office of Science User Facility under the same contract. Additional support for DESI was provided by the U.S. National Science Foundation (NSF), Division of Astronomical Sciences under Contract No. AST-0950945 to the NSF’s National Optical-Infrared Astronomy Research Laboratory; the Science and Technology Facilities Council of the United Kingdom; the Gordon and Betty Moore Foundation; the Heising-Simons Foundation; the French Alternative Energies and Atomic Energy Commission (CEA); the National Council of Humanities, Science and Technology of Mexico (CONAHCYT); the Ministry of Science, Innovation and Universities of Spain (MICIU/AEI/10.13039/501100011033), and by the DESI Member Institutions: \url{https://www.desi.lbl.gov/collaborating-institutions}. Any opinions, findings, and conclusions or recommendations expressed in this material are those of the author(s) and do not necessarily reflect the views of the U. S. National Science Foundation, the U. S. Department of Energy, or any of the listed funding agencies.

The authors are honored to be permitted to conduct scientific research on I'oligam Du'ag (Kitt Peak), a mountain with particular significance to the Tohono O’odham Nation.

% Bibliography
\bibliographystyle{JHEP}
\bibliography{references.bib}

\appendix
\section{Potential Feasibility with LyCAN}\label{sec:appendix_lycan}

The \lya Continuum Analysis Network \citep[LyCAN;][]{turner_2024} is a convolutional neural network that predicts the unabsorbed quasar continuum between $1040-1600\,\angstrom$ strictly based on the longer-wavelength region ($1216-1600\,\angstrom$). Since the network predicts the unabsorbed continuum independently of the forest region, it retains large-scale information and may eliminate the need for the distortion matrix. LyCAN was trained on a combination of DESI mock spectra representative of the five-year survey and synthetic spectra generated from low-z Hubble Space Telescope (HST)/Cosmic Origins Spectrograph (COS) observations. It achieves excellent performance on spectra from these datasets reserved for testing purposes ($\lesssim2\%$ median error in the forest region).

\begin{figure*}
    \centering
    \includegraphics[width=\linewidth]{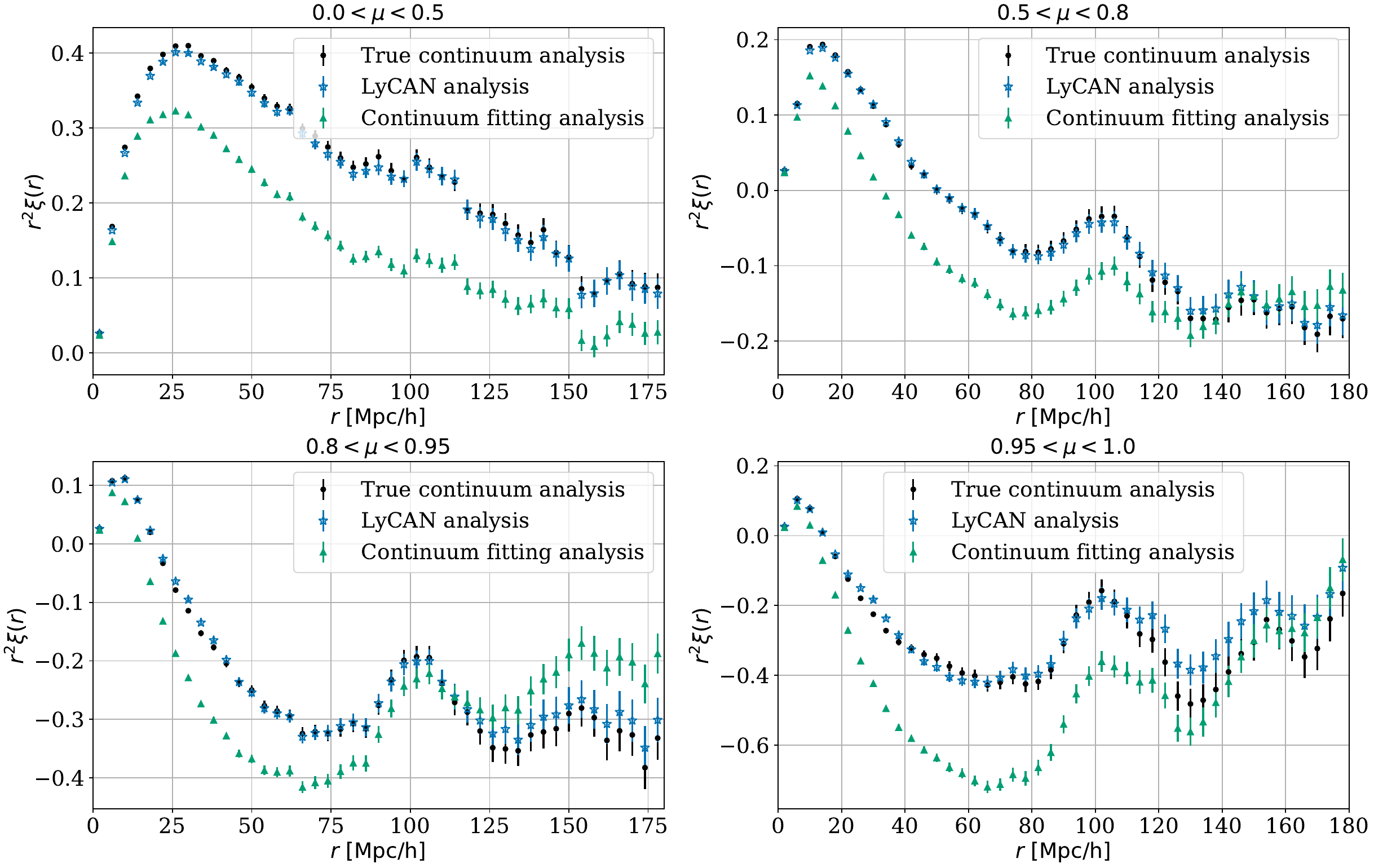}
    \caption{Three measurements of the \lya autocorrelation function $\xi(r)$ in four wedges of $\mu=r_\parallel/r$ on an uncontaminated DESI mock representative of the full five-year survey: the measurement from the standard continuum fitting approach (\textit{teal}), the measurement from the true continuum analysis (\textit{black}), and the measurement from LyCAN-predicted continua (\textit{blue}). These measurements are shown for illustrative purposes, demonstrating the future feasibility of a LyCAN-based full-shape analysis.}
    \label{fig:lycan-autocorr}
\end{figure*}

We have made several updates to LyCAN since its introduction in \cite{turner_2024}. While the original network demonstrated strong overall performance in forest continuum prediction, it introduced some spurious correlations in the 3D \lya correlation functions relative to the true continuum case. We found that this was due to the addition of systematic perturbations in the synthetic spectra generated from high-resolution HST/COS data to dampen emission line strengths (see Section 3.2 of \cite{turner_2024}). After modifying this perturbation scheme to be centered around zero, we re-optimized and re-trained LyCAN using the procedures described in Section 4.1 of \cite{turner_2024}. This updated version of LyCAN yields an undistorted measurement of the correlation function when applied to a DESI mock reserved for testing that is representative of the five-year survey, as illustrated in Figure~\ref{fig:lycan-autocorr}. This result motivates this work, as a continuum prediction method that produces undistorted correlation function measurements may soon be within reach. The revised neural network architecture is presented in Table~\ref{tab:arch}.

\begin{table}
	\centering
	\caption{Updated LyCAN architecture. The first column specifies the layer type. The \textit{size} column specifies the number of filters for convolutional layers or neurons for dense layers. In the case of the max pooling layers, pixels are pooled in groups of two. We use the linear activation function for the output layer, and the ReLU activation function for all other layers where applicable. The number of nodes in the output layer is equal to the number of pixels in the predicted continuum. The kernel size is six for the convolutional layers.}
	\label{tab:arch}
	\begin{tabular}{lcr}
		\hline
		Layer type & Size & Activation function\\
		\hline
		Conv1D (input) & 64 & ReLU\\
        MaxPooling1D & 2 & \\
		Conv1D & 16 & ReLU\\
        MaxPooling1D & 2 & \\
        Conv1D & 48 & ReLU\\
        MaxPooling1D & 2 & \\
        Conv1D & 32 & ReLU\\
        MaxPooling1D & 2 & \\
        Flatten &  & \\
		Dense & 96 & ReLU\\
        Dense & 160 & ReLU\\
        Dense & 224 & ReLU\\
        Dense & 256 & ReLU\\
        Dense (output) & 699 & Linear\\
		\hline
	\end{tabular}
\end{table}

LyCAN continuum predictions also show slight biases with respect to signal-to-noise ratio (SNR) and redshift, under-predicting the true continuum in low-SNR or low-redshift spectra and over-predicting it at high redshift \citep{turner_2024}. While this is a small effect ($\sim2\%$, as seen in Figure~12 of \cite{turner_2024}), it may still warrant correction or further network optimization before LyCAN can be reliably applied to measure undistorted correlation functions. Ongoing efforts aim to improve LyCAN and mitigate these biases in order to achieve accurate correlation function measurements across a wide range of mocks. We therefore do not use LyCAN continuum predictions in our forecasts and instead focus on the idealized case of the true continuum.

\section{Validation of the Hartlap Correction}\label{sec:appendix_hartlap}

When a covariance matrix is estimated from a limited number of samples relative to the size of the data vector, its inverse is a biased estimate of the true inverse covariance \citep[e.g.][]{hartlap_2007,percival_covariance_2014,percival_2022}. If left uncorrected, this bias leads to underestimated uncertainties on inferred parameters. A common mitigation is the Hartlap correction \citep{hartlap_2007}, which rescales the inverse covariance matrix during likelihood evaluation:
\begin{equation}
h = \frac{N-p-1}{N},
\end{equation}
where $N$ is the number of independent samples used to estimate the covariance matrix and $p$ is the size of the data vector.

\begin{figure*}
    \centering
    \includegraphics[width=\linewidth]{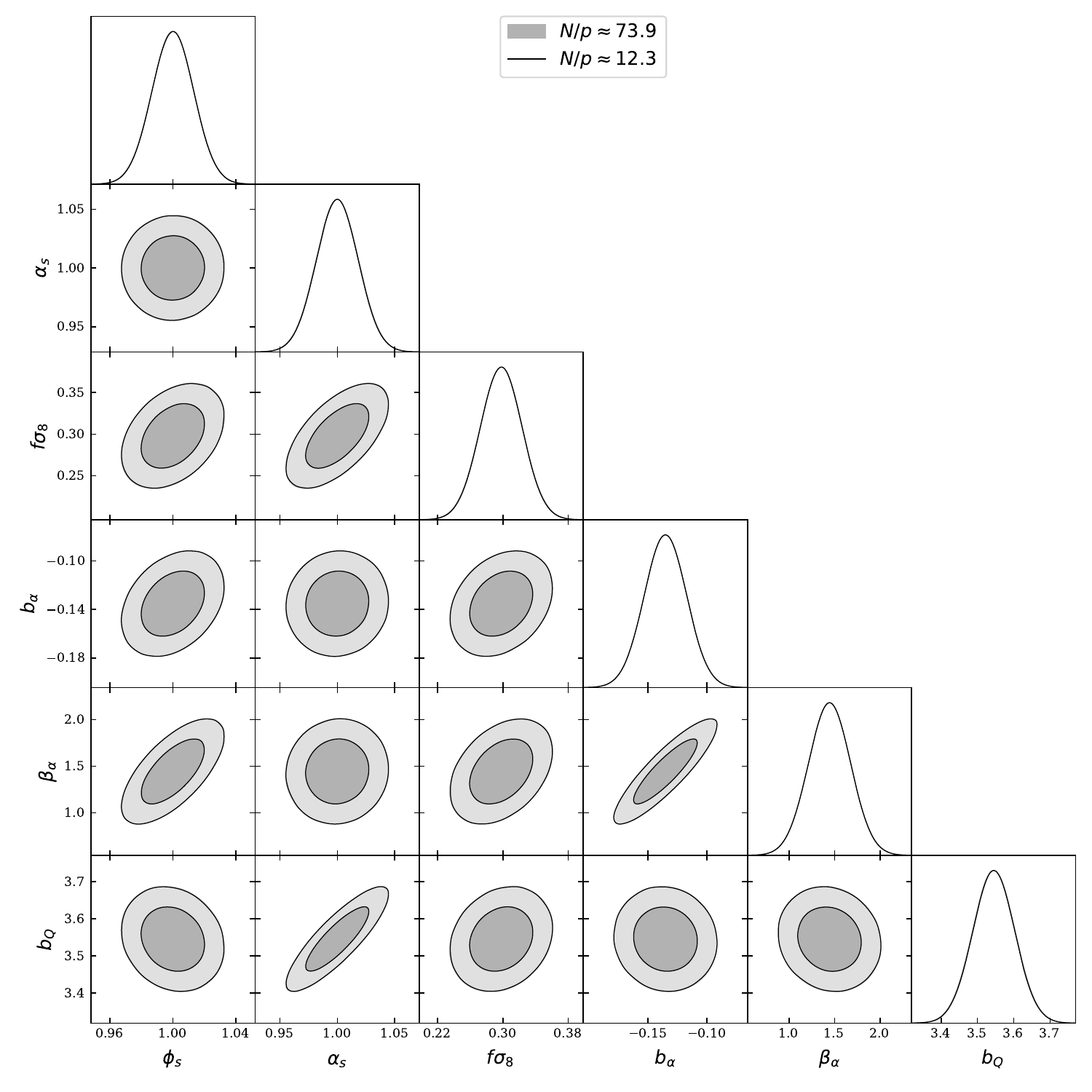}
    \caption{Forecasts using two covariance matrices from fully contaminated \texttt{NSIDE=16} mocks based on stacks of 300 ($N/p\approx73.9$, \textit{solid gray contours}) or 50 ($N/p\approx12.3$, \textit{black contours}). Appropriate Hartlap corrections are applied in each case. The results are indistinguishable, validating the correction in the lower $N/p$ regime.}
    \label{fig:hartlap_validation_contour}
\end{figure*}

However, the Hartlap correction is only an approximate solution, and its accuracy degrades as $N$ approaches $p$. To validate its reliability in our analysis, we test its performance across a range of $N/p$ values. We first compute a reference covariance matrix from 300 fully contaminated mocks in the baseline continuum fitting analysis, which includes separations from $30-180\,h^{-1}\,\mathrm{Mpc}$ in \texttt{NSIDE}=16 HEALPix pixels. This setup corresponds to $N=344,040$ measurements and a data vector of size $p=4653$, yielding $N/p\approx74$. In this regime, the Hartlap correction is negligible and the covariance is expected to be robust. We then compute the same covariance using only 50 fully contaminated mocks, matching the number of realizations available for our uncontaminated analysis. We apply the corresponding Hartlap correction in both cases and perform maximum likelihood analysis. As shown in Figure~\ref{fig:hartlap_validation_contour}, the resulting constraints are indistinguishable, confirming that the Hartlap-corrected inverse covariance is reliable at $N/p\approx12$.

\begin{figure}
    \centering
    \includegraphics[width=0.7\linewidth]{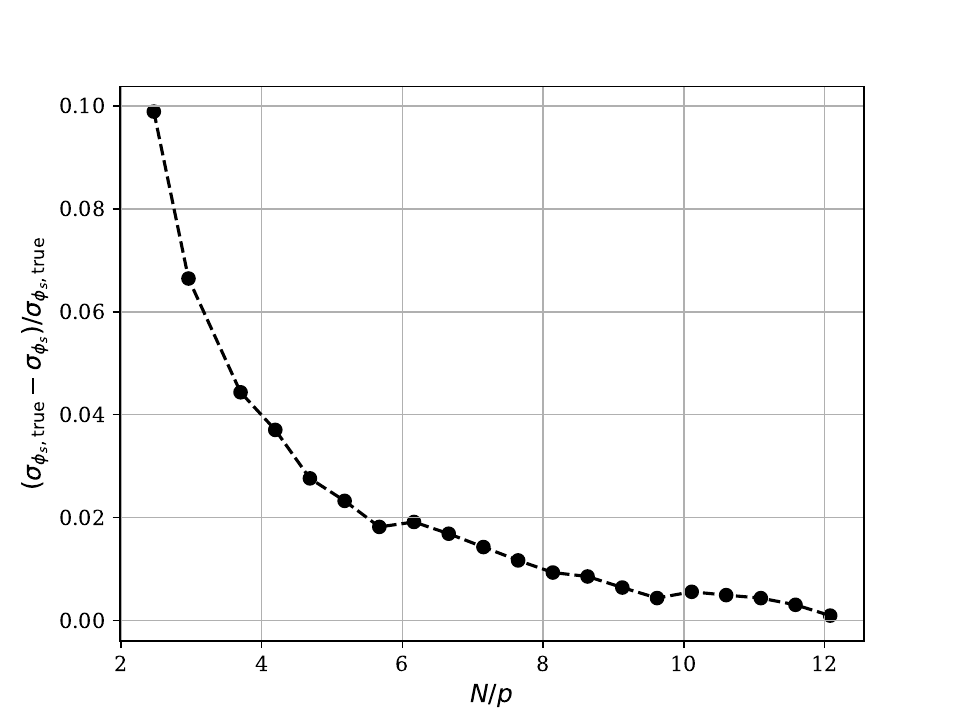}
    \caption{Fractional error in the inferred uncertainty on $\phi_s$ in different $N/p$ regimes using Hartlap-corrected inverse covariances. We require $N/p\gtrsim5$ to limit our error on inferred parameter uncertainties to within $\sim2\%$.}
    \label{fig:hartlap_error}
\end{figure}

To test the limits of this approximation, we compute additional covariance matrices from the uncontaminated mocks by varying the number of realizations and examining the resulting forecasts for $\phi_s$. For each value of $N$, we apply the Hartlap correction and evaluate the fractional error in the inferred uncertainty on $\phi_s$ relative to the result from the full 50-mock covariance (our reference ``truth"). The results are shown in Figure~\ref{fig:hartlap_error}. We find that achieving $N/p\gtrsim5$ is sufficient to limit errors in parameter uncertainties to within $\sim2\%$. This criterion motivates our use of covariances computed from \texttt{NSIDE}=14 HEALPix pixels up to $240\,h^{-1}\,\mathrm{Mpc}$ in the main analysis.

We note that the Hartlap factor does not account for the additional uncertainty arising from the sample covariance matrix being a single realization of an underlying distribution, which propagates to increased parameter uncertainties \citep{dodelson_schneider_2013}. This results in a non-Gaussian likelihood \citep{sellentin_heavens_2016}, but can be approximated using a Gaussian likelihood and additional correction factors \citep[e.g.][]{dodelson_schneider_2013,percival_covariance_2014,percival_2022}. We do not apply these additional corrections since our validation tests demonstrate that the Hartlap correction alone can sufficiently recover the truth in our analysis.

\section{Author Affiliations}
\label{sec:affiliations}

\noindent \hangindent=.5cm $^{a}${Center for Cosmology and AstroParticle Physics, The Ohio State University, 191 West Woodruff Avenue, Columbus, OH 43210, USA}

\noindent \hangindent=.5cm $^{b}${Department of Astronomy, The Ohio State University, 4055 McPherson Laboratory, 140 W 18th Avenue, Columbus, OH 43210, USA}

\noindent \hangindent=.5cm $^{c}${Lawrence Berkeley National Laboratory, 1 Cyclotron Road, Berkeley, CA 94720, USA}

\noindent \hangindent=.5cm $^{d}${Department of Physics, The Ohio State University, 191 West Woodruff Avenue, Columbus, OH 43210, USA}

\noindent \hangindent=.5cm $^{e}${Department of Physics, Boston University, 590 Commonwealth Avenue, Boston, MA 02215 USA}

\noindent \hangindent=.5cm $^{f}${Dipartimento di Fisica ``Aldo Pontremoli'', Universit\`a degli Studi di Milano, Via Celoria 16, I-20133 Milano, Italy}

\noindent \hangindent=.5cm $^{g}${INAF-Osservatorio Astronomico di Brera, Via Brera 28, 20122 Milano, Italy}

\noindent \hangindent=.5cm $^{h}${Department of Physics \& Astronomy, University College London, Gower Street, London, WC1E 6BT, UK}

\noindent \hangindent=.5cm $^{i}${Institut de F\'{i}sica d'Altes Energies (IFAE), The Barcelona Institute of Science and Technology, Edifici Cn, Campus UAB, 08193, Bellaterra (Barcelona), Spain}

\noindent \hangindent=.5cm $^{j}${Instituto de F\'{\i}sica, Universidad Nacional Aut\'{o}noma de M\'{e}xico,  Circuito de la Investigaci\'{o}n Cient\'{\i}fica, Ciudad Universitaria, Cd. de M\'{e}xico  C.~P.~04510,  M\'{e}xico}

\noindent \hangindent=.5cm $^{k}${Department of Astronomy \& Astrophysics, University of Toronto, Toronto, ON M5S 3H4, Canada}

\noindent \hangindent=.5cm $^{l}${Department of Physics \& Astronomy and Pittsburgh Particle Physics, Astrophysics, and Cosmology Center (PITT PACC), University of Pittsburgh, 3941 O'Hara Street, Pittsburgh, PA 15260, USA}

\noindent \hangindent=.5cm $^{m}${University of California, Berkeley, 110 Sproul Hall \#5800 Berkeley, CA 94720, USA}

\noindent \hangindent=.5cm $^{n}${Departamento de F\'isica, Universidad de los Andes, Cra. 1 No. 18A-10, Edificio Ip, CP 111711, Bogot\'a, Colombia}

\noindent \hangindent=.5cm $^{o}${Observatorio Astron\'omico, Universidad de los Andes, Cra. 1 No. 18A-10, Edificio H, CP 111711 Bogot\'a, Colombia}

\noindent \hangindent=.5cm $^{p}${Institut d'Estudis Espacials de Catalunya (IEEC), c/ Esteve Terradas 1, Edifici RDIT, Campus PMT-UPC, 08860 Castelldefels, Spain}

\noindent \hangindent=.5cm $^{q}${Institute of Cosmology and Gravitation, University of Portsmouth, Dennis Sciama Building, Portsmouth, PO1 3FX, UK}

\noindent \hangindent=.5cm $^{r}${Institute of Space Sciences, ICE-CSIC, Campus UAB, Carrer de Can Magrans s/n, 08913 Bellaterra, Barcelona, Spain}

\noindent \hangindent=.5cm $^{s}${University of Virginia, Department of Astronomy, Charlottesville, VA 22904, USA}

\noindent \hangindent=.5cm $^{t}${Fermi National Accelerator Laboratory, PO Box 500, Batavia, IL 60510, USA}

\noindent \hangindent=.5cm $^{u}${Institut d'Astrophysique de Paris. 98 bis boulevard Arago. 75014 Paris, France}

\noindent \hangindent=.5cm $^{v}${IRFU, CEA, Universit\'{e} Paris-Saclay, F-91191 Gif-sur-Yvette, France}

\noindent \hangindent=.5cm $^{w}${The Ohio State University, Columbus, 43210 OH, USA}

\noindent \hangindent=.5cm $^{x}${Department of Physics, The University of Texas at Dallas, 800 W. Campbell Rd., Richardson, TX 75080, USA}

\noindent \hangindent=.5cm $^{y}${NSF NOIRLab, 950 N. Cherry Ave., Tucson, AZ 85719, USA}

\noindent \hangindent=.5cm $^{z}${Department of Physics, Southern Methodist University, 3215 Daniel Avenue, Dallas, TX 75275, USA}

\noindent \hangindent=.5cm $^{aa}${Department of Physics and Astronomy, University of California, Irvine, 92697, USA}

\noindent \hangindent=.5cm $^{ab}${Sorbonne Universit\'{e}, CNRS/IN2P3, Laboratoire de Physique Nucl\'{e}aire et de Hautes Energies (LPNHE), FR-75005 Paris, France}

\noindent \hangindent=.5cm $^{ac}${Departament de F\'{i}sica, Serra H\'{u}nter, Universitat Aut\`{o}noma de Barcelona, 08193 Bellaterra (Barcelona), Spain}

\noindent \hangindent=.5cm $^{ad}${Instituci\'{o} Catalana de Recerca i Estudis Avan\c{c}ats, Passeig de Llu\'{\i}s Companys, 23, 08010 Barcelona, Spain}

\noindent \hangindent=.5cm $^{ae}${Departamento de F\'{\i}sica, DCI-Campus Le\'{o}n, Universidad de Guanajuato, Loma del Bosque 103, Le\'{o}n, Guanajuato C.~P.~37150, M\'{e}xico}

\noindent \hangindent=.5cm $^{af}${Instituto Avanzado de Cosmolog\'{\i}a A.~C., San Marcos 11 - Atenas 202. Magdalena Contreras. Ciudad de M\'{e}xico C.~P.~10720, M\'{e}xico}

\noindent \hangindent=.5cm $^{ag}${Department of Physics and Astronomy, University of Waterloo, 200 University Ave W, Waterloo, ON N2L 3G1, Canada}

\noindent \hangindent=.5cm $^{ah}${Perimeter Institute for Theoretical Physics, 31 Caroline St. North, Waterloo, ON N2L 2Y5, Canada}

\noindent \hangindent=.5cm $^{ai}${Waterloo Centre for Astrophysics, University of Waterloo, 200 University Ave W, Waterloo, ON N2L 3G1, Canada}

\noindent \hangindent=.5cm $^{aj}${Space Sciences Laboratory, University of California, Berkeley, 7 Gauss Way, Berkeley, CA  94720, USA}

\noindent \hangindent=.5cm $^{ak}${Instituto de Astrof\'{i}sica de Andaluc\'{i}a (CSIC), Glorieta de la Astronom\'{i}a, s/n, E-18008 Granada, Spain}

\noindent \hangindent=.5cm $^{al}${Department of Physics and Astronomy, Sejong University, 209 Neungdong-ro, Gwangjin-gu, Seoul 05006, Republic of Korea}

\noindent \hangindent=.5cm $^{am}${CIEMAT, Avenida Complutense 40, E-28040 Madrid, Spain}

\noindent \hangindent=.5cm $^{an}${Department of Physics, University of Michigan, 450 Church Street, Ann Arbor, MI 48109, USA}

\noindent \hangindent=.5cm $^{ao}${University of Michigan, 500 S. State Street, Ann Arbor, MI 48109, USA}

\noindent \hangindent=.5cm $^{ap}${Department of Physics \& Astronomy, Ohio University, 139 University Terrace, Athens, OH 45701, USA}

\noindent \hangindent=.5cm $^{aq}${Excellence Cluster ORIGINS, Boltzmannstrasse 2, D-85748 Garching, Germany}

\noindent \hangindent=.5cm $^{ar}${University Observatory, Faculty of Physics, Ludwig-Maximilians-Universit\"{a}t, Scheinerstr. 1, 81677 M\"{u}nchen, Germany}

\noindent \hangindent=.5cm $^{as}${National Astronomical Observatories, Chinese Academy of Sciences, A20 Datun Road, Chaoyang District, Beijing, 100101, P.~R.~China}

\end{document}